%% file: ME.tex
\documentclass[graybox]{svmult}

\usepackage{mathptmx}       
\usepackage{epsfig,amsmath,amssymb,graphics} 
\usepackage{helvet}         
\usepackage{courier}        
\usepackage{type1cm}        
\usepackage{makeidx}         
\usepackage{graphicx}        
\usepackage{multicol}        
\usepackage[bottom]{footmisc}

\makeindex             

\begin{document}

\title*{Fractional Maps as Maps with Power-Law Memory}
\author{Mark Edelman}
\institute{Mark Edelman \at Stern College at Yeshiva University, 
245 Lexington Street, New York, NY 10016 and 
\at Courant Institute of Mathematical Sciences at NYU, New York,
  NY 10012 
\at \email{edelman@cims.nyu.edu}}
%
%
\maketitle

\abstract{The study of systems with memory requires methods which are
different from the methods used in regular dynamics. Systems
with  power-law memory in many cases can be described by  fractional
differential equations, which are integro-differential equations.
To study the general properties of  nonlinear fractional dynamical 
systems  we use fractional maps, which are discrete nonlinear systems with 
power-law memory derived from fractional differential equations. 
To study fractional maps we use the notion of  $\alpha$-families of 
maps depending on a 
single parameter $\alpha > 0$ which is the order of the fractional
derivative in a nonlinear fractional differential equation describing 
a system experiencing periodic kicks.  $\alpha$-families of maps
represent a very general form of  multi-dimensional nonlinear maps with 
power-law memory, in which the weight of the previous 
state at time $t_i$ in defining the present state at time $t$ is proportional
to $(t-{t_i})^{\alpha-1}$.
They may be applicable to studying some systems with
memory such as viscoelastic materials, electromagnetic fields in dielectric
media, Hamiltonian systems, adaptation in biological systems, human
memory, etc. 
Using the fractional logistic 
and standard $\alpha$-families of maps 
as examples
we demonstrate that the phase space of 
nonlinear fractional 
dynamical systems may contain  periodic sinks, 
attracting slow diverging trajectories,
attracting accelerator mode trajectories, chaotic attractors, and cascade
of bifurcations type trajectories 
whose properties are different from properties of attractors in regular
dynamical systems.
}

\section{Introduction}
\label{sec:1}
Many natural and social systems are systems with memory. Their
mathematical description requires solving integro-differential
equations and is quite complicated. Maps with memory are used to model
real systems with memory in order to derive their basic properties.
 
\subsection{Systems with Memory}
\label{sec:1.1}

Writing this text I am recalling the content of my latest papers. It is easy
to recall the content of my last paper but it becomes  more and more difficult
as I try to recall 
papers that are more and more distant in time. 
Memory is a
significant property of human beings and is the subject of  extensive
biophysical and psychological research. As it has been demonstrated in
experiments, forgetting - the accuracy on a memory tasks decays 
as a power law,  $\sim
t^{-\beta}$, with $0<\beta<1$ 
\cite{Kahana,Rubin,Wixted1,Wixted2,Adaptation1}. 
It is interesting that fractional maps
corresponding to fractional differential equations of the order
$0<\alpha<1$  are maps with the power-law decaying memory in which the power is
$-\beta=\alpha-1$ and $0<\beta<1$ \cite{DNC}. 
Human learning is closely related to memory.  It also can be
described by a power law: the reduction in reaction times that comes with
practice is a power function of the number of training trials \cite{Anderson}.
There are multiple publications where power-law adaptation has been applied 
in describing the dynamics of biological systems at levels ranging from 
single ion channels up to human psychophysics 
\cite{Adaptation3,Adaptation4,Adaptation2,Adaptation5,Adaptation1,Adaptation6}.

Power-law memory applies not only to the human being  
as a whole, but also to the hierarchy of its building blocks, from
individual neurons and proteins to the
tissue of individual organs. 
It has been shown recently 
\cite{Lund2,Lund1} that processing of external stimuli by individual neurons
can be described by fractional differentiation. The orders of 
fractional derivatives $\alpha$ derived for different types of neurons fall 
within the interval $[0,1]$. 
For neocortical pyramidal neurons it is quite small: $\alpha \approx 0.15$.
Fluctuations within 
single protein molecules demonstrate a power-law memory kernel with the
exponent $-0.51 \pm 0.07$ \cite{Protein}.
 
Viscoelastic properties of human tissues were demonstrated in many
examples: the brain and the central nervous system in general
\cite{TissueNerv,TissueBrain2,TissueBrain1}, the breast \cite{Coussot},
the liver \cite{TissueLiver2,TissueLiver1}, the spleen \cite{TissueSpleen},
the prostate \cite{TissueProstate1,TissueProstate2}, 
the arteries \cite{TissueArteries1,TissueArteries2},
the muscles  \cite{TissueMuscle} (see also references for some other human 
and animal organs
tissues \cite{DFCV2005,Magin,MGE2010,TissueBov,Frog2008}).
Viscoelastic materials obey the following stress-strain relationship:
\begin{equation}
\sigma(t)=E(\gamma)\frac{d^\alpha\gamma(t)}{dt^\alpha},
\label{VE}
\end{equation}
where $\sigma$ is the stress, $\gamma$ is the strain,
$\alpha$ is the order of the fractional derivative, and $t$ is time. 
In most of the cases for human tissues $0<\alpha<1$ and is close to
zero. In some cases, e.g. for modeling of the accurate placement of the 
needle tip into the target tissue during needle insertion treatments for
liver tumors, nonlinearity of $E(\gamma)$ should be taken into account 
\cite{TissueLiver2}. In the last example a simple quadratic nonlinearity
and $\alpha=0.1$ were used. 

A Fourier transform of a  fractional derivative is \cite{KST,Podlubny,SKM}
\begin{equation}
F\{D^{\alpha}g(t);\omega \}=(-i \omega)^{\alpha}\hat{g}(\omega),
\label{Fourier}
\end{equation}
where $\hat{g}(\omega)=F\{g(t);\omega \}$. As a result, whenever 
the term  $(\omega)^{\alpha}\hat{g}(\omega)$ appears in the frequency
domain, there is a good chance that function $g(t)$ is a solution of a
fractional differential equation with a fractional derivative of the
order $\alpha$ and the corresponding system is a system with power-law
memory. Well known examples of such systems are dielectrics.   
Electromagnetic fields in dielectric media are described by equations with
time fractional derivatives due to the 'universal' response - the power-law
frequency dependence of the dielectric susceptibility in a wide range of
frequencies \cite{TDia2008a,TDia2008b,TDia2009,TarBook}. Similarly,
elastic wave attenuation in biological tissue over a wide range of
frequencies follows the power law $\alpha(\omega) \propto \omega^{\eta}$
with $\eta \in [0,2]$
\cite{TissueWaves1,TissueWaves2,TissueWaves3,TissueWaves4} which implies 
a fractional wave equation. The establishment of accurate fractional 
wave-propagation models is important for many medical applications 
\cite{TissueWaves3}.

Above we concentrated on  biological systems with memory in order
to emphasize the importance of the study of nonlinear fractional
dynamical systems described by fractional differential equations 
of the order $0<\alpha<2$ and especially $\alpha$ close to zero
which is a major subject of the following sections 
(Sec.~\ref{LT1}~and~\ref{BN12}).
Now we'll list some (not all) 
other examples of systems with power-law memory. 
As  has been mentioned above, time fractional derivatives and
correspondingly systems with power-law memory in many cases are used to
describe viscoelasticity and rheology (for the original papers and reviews
see \cite{Visc1,Visc2,Visc3,Visc4,Visc5,Visc6,Visc7,Visc8}, 
for nonlinear effects see \cite{NLVisc1,NLVisc2,NLVisc3,NLVisc4}).
Electromagnetic fields in dielectric media were also mentioned above.
Hamiltonian systems and billiards are also 
systems with power-law memory, in which the
fractal structure of the phase space and
stickiness of trajectories in time imply description of transport by
the fractional (factional time and space derivatives) Fokker-Plank-Kolmogorov
equation \cite{ZasBook,ZE2000,ZE2004,ZE1997}.
In some cases \cite{KBT1,KBT2,KST} fractional differential equations are equivalent 
to the Volterra integral equations of the second kind.
Systems considered in population biology and epidemiology are systems with 
memory and Volterra integral equations  are frequently used to describe such
systems \cite{PopBioBook2001,HoppBook1975}. 
Long-term memory provides more robust control in liner and nonlinear 
control theory (see books \cite{Control1,ControlBook2011}).

\subsection{Maps with Memory}
\label{sec:1.2}
As in the study of regular dynamics, in the study 
of systems with memory use of discrete maps significantly simplifies 
investigation of the general properties of the corresponding systems. 
In some cases of kicked systems maps are equivalent to the 
original differential equations. Historically, maps with memory 
were first considered as analogues of the integro-differential
equations of non-equilibrium statistical physics \cite{MM1,MM2,MM6},
with regards to thermodynamic theory of systems with memory 
\cite{MM5}, and to model non-Markovian processes in general 
\cite{MM3,MM4}.
The most general form  of a map with memory is
\begin{equation}
\mathbf{x}_{n+1}=\mathbf{f}_{n+1}(\mathbf{x}_n,\mathbf{x}_{n-1},...,\mathbf{x}_0,P),
\label{MMGeneral}
\end{equation} 
where $\bf{x}_k$ are $N$-dimensional vectors, $k,N \in  \mathbb{Z}$, $k \ge 0$, 
and $P$ is a set of parameters.
It is almost impossible to derive the general properties of systems with
memory from Eq.~(\ref{MMGeneral}) and simplified forms of maps with memory are
used. The most commonly used form is the one-dimensional  map 
with long-term memory
\begin{equation}
x_{n+1}=\sum^{n}_{k=0}V_{\alpha}(n,k)G_K(x_k),
\label{LTM}
\end{equation} 
where $V_{\alpha}(n,k)$ and $\alpha$ characterize memory effects
and $K$ is a parameter.
In many cases weights are taken as convolutions with
$V_{\alpha}(n,k)=V_{\alpha}(n-k)$. The particular form of 
Eq.~(\ref{LTM}) with
constant weights
\begin{equation}
x_{n+1}=c\sum^{n}_{k=0}G_K(x_k)
\label{Full}
\end{equation} 
is called a full-memory map. It is easy to note that Eq.~(\ref{Full})
is equivalent to 
\begin{equation}
x_{n+1}=x_n+cG_K(x_n),
\label{F1Step}
\end{equation} 
which means that maps with  full memory are maps with 
one-step memory in which all memory is accumulated in the present
state of a system and the next values of map variables
are fully defined by their present values. 
We won't consider maps with short memory in which the number of
terms in the sum in  Eq.~(\ref{LTM}) is bounded (from $k=n-M+1$ 
to $k=n$).

Initial investigations of long-term memory maps were done 
mostly on  different modifications of the logistic map 
and exponential memory. The general applicability
of their results to systems with memory in general is limited.  
Recently Stanislavsky \cite{MM7} considered the maps Eq.~(\ref{LTM})
with $G_K(x)=Kx(1-x)$ (the logistic map) and the weights $V_{\alpha}(n,k)$ 
as a combination
of power-law functions taken from one of algorithms of numerical
fractional integration.   
He came to the conclusion that increase in  long-term memory effects 
leads to a less chaotic behavior.

First maps with power-law memory equivalent to fractional differential 
equations were derived in \cite{MM9,MM8,TarBook,MM10,MM11} by integrating 
fractional differential equations describing systems under
periodic kicks. The method used is similar to the way in which 
the universal map is derived in regular dynamics.

\subsection{Universal Map}
\label{sec:1.3}

In the following section (Sec.~\ref{UFM}) we will modify the way
presented in Sec.~\ref{sec:1.3}
to derive the universal map
in regular dynamics (see \cite{Chir}, and Ch. 5 from \cite{ZasBook})
in order to derive the universal fractional map.  

The universal map can be derived from the differential equation
\begin{equation}
\ddot{x}+KG(x) \sum^{\infty}_{n=-\infty} \delta \Bigl(\frac{t}{T}-(n+\varepsilon)
\Bigr)=0,
\label{UMDE}
\end{equation}
where $0 < \varepsilon < 1$ and $K$ is a parameter,
with the initial conditions:
\begin{equation}
x(0)=x_0, \    \ p(0)=\dot{x}(0)=p_0.
\label{SMDEIC}
\end{equation}
This equation is equivalent to the Volterra
integral equation of second kind
\begin{equation}
x(t)=x_0 + p_0t - K\int^{t}_0 d \tau  G(x( \tau )) \sum^{\infty}_{n=-\infty}
\delta \Bigl(\frac{\tau}{T}-(n+\varepsilon)\Bigr)( t-\tau ).
\label{Volt2D}
\end{equation}
Eq.~(\ref{Volt2D}) for $(n+\varepsilon)T<t<(n+1+\varepsilon)T$ has a solution
\begin{eqnarray}
&&x(t)=x_0 + p_0t -  
KT\sum^{n}_{k=0} G(x( Tk+T\varepsilon)) ( t-Tk-T\varepsilon),  \nonumber\\
&&p(t)=\dot{x}(t)= p_0 -  
KT\sum^{n}_{k=0} G(x( Tk+T\varepsilon)).
\label{Volt2Dxp}
\end{eqnarray}
After the introduction of the map variables
\begin{equation}
x_{n}=x(Tn), \   \ p_{n}=p(Tn)
\label{xnpn}
\end{equation}
Eq.~(\ref{Volt2Dxp}) considered for time instances t=(n+1)T gives
\begin{eqnarray}
&&x_{n+1}=x_0 + p_0(n+1)T -   
KT^2\sum^{n}_{k=0} G(x( Tk+T\varepsilon)) (n-k+1-\varepsilon),  \nonumber\\
&&p_{n+1}= p_0 -  
KT\sum^{n}_{k=0} G(x( Tk+T\varepsilon)).
\label{Volt2Dxpn}
\end{eqnarray}
As it follows from Eq.~(\ref{Volt2Dxp}),  $\dot{x}(t)=p(t)$ is a
bounded function with the discontinuities at the time instances of the kicks
(at $t=Tk+T\varepsilon$) and $x(t)$ is a continuous function.
This allows us to calculate $G(x)$ at the time
instances of the kicks. In the limit   
$\varepsilon  \rightarrow 0$ Eq.~(\ref{Volt2Dxpn}) gives 
\begin{eqnarray}
&&x_{n+1}=x_0 + p_0(n+1)T -   
KT^2\sum^{n}_{k=0} G(x_k) ( n-k+1),   \nonumber\\ 
&&p_{n+1}= p_0 -  
KT\sum^{n}_{k=0} G(x_k).
\label{Umapxp}
\end{eqnarray}
Eq.~(\ref{Umapxp}) is a form of the universal map which allows further
simplifications.  
It can be written in a symmetric form as a map with full memory 
(see Sec.~\ref{sec:1.2}):
\begin{eqnarray}
&&x_{n+1}= x_0 +  
T\sum^{n+1}_{k=1} p_k, \nonumber\\ 
&&p_{n+1}= p_0 -  
KT\sum^{n}_{k=0} G(x_k).
\label{UmappSimxp}
\end{eqnarray}
As we saw in  Sec.~\ref{sec:1.2}, maps with full memory are equivalent to
maps with one-step memory.
Map Eq.~(\ref{UmappSimxp}) can be written
as the iterative area preserving
($\partial(p_{n+1},x_{n+1})/\partial(p_{n},x_{n})=1$) process with
one-step memory which is called the universal map:
\begin{equation}
p_{n+1}= p_{n} - KTG(x_n),
\label{UMp}
\end{equation}
\begin{equation}
x_{n+1}= x_{n}+ p_{n+1}T.
\label{UMx}
\end{equation}
This map represents the relationship between the values of the physical
variables in Eq.~(\ref{UMDE}) on the left sides of the consecutive kicks.
The standard map may be obtained from the universal map by assuming
$G(x)=\sin(x)$:
\begin{eqnarray}
&&p_{n+1}= p_{n} - K \sin x,  \ \ \ \ ({\rm mod} \ 2\pi ),  
\nonumber \\
&&x_{n+1}= x_{n}+ p_{n+1},  \ \ \ \ ({\rm mod} \ 2\pi ).
\label{SM}
\end{eqnarray}
Here we assumed $T=1$ and consider this map on a torus (${\rm mod} \ 2\pi$).

Derivation of the fractional universal map in the next section
(Sec.~\ref{UFM}) follows \cite{DNC} and the analysis of this map 
for $\alpha \in (0,1)$ and  $\alpha \in (1,2)$ in 
Secs.~\ref{LT1}~and~\ref{BN12} follows \cite{ME1,DNC,ME3,ME4}.

\section{Fractional Universal Map}
\label{UFM}

The one-dimensional logistic map
\begin{equation}
x_{n+1}=K x_{n}(1-x_{n})
\label{LM}
\end{equation}
may be presented in the 2D form
\begin{eqnarray}
&&p_{n+1}= - G_{lK}(x_n),   \nonumber\\   
&&x_{n+1}= x_{n}+ p_{n+1},
\label{LM2Dn}
\end{eqnarray}
where
\begin{equation}
G_{lK}(x)=x-Kx(1-x).
\label{GLM}
\end{equation}
It can't be written as a particular form of the universal map Eqs.~(\ref{UMp})
and  (\ref{UMx}).
In order to derive the logistic map from the universal map we'll introduce the notion of the n-dimensional universal map depending 
on a single parameter.

\subsection{Universal Integer-Dimensional Maps}
\label{UnM}

\begin{figure}[b]
\sidecaption
\includegraphics[scale=.35]{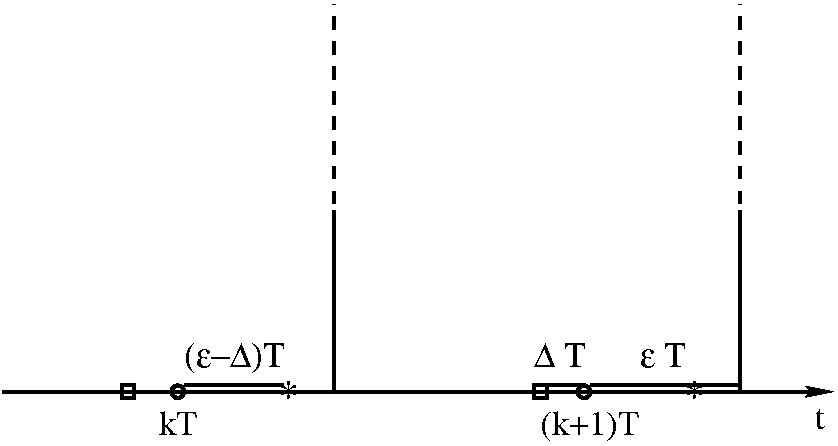}
\caption{The universal map is a relationship between values of $x(t)$
  considered at the times $kT$ (small circles). The kicks occur at the
  time instances $(k+\varepsilon)T$ (vertical lines). With the time delay 
  $\Delta  T$ (distance between the squares and the circles) the function
  $G_K(x(t))$ is calculated at the time instances $t=(k+\varepsilon-\Delta)T$ (stars).
  }
\label{kicks}      
\end{figure}

Solution of the one-dimensional analog of  Eq.~(\ref{UMDE}) would require 
calculations of the function $G(x)$ at the time instances of 
the kicks $T(n+\varepsilon)$ 
at which $x(t)$ is
discontinuous. To enable us to introduce the universal fractional map
we'll include a time delay $\Delta T$ into the argument of the function
$G(x(t))$ (see Fig.~\ref{kicks}).
In order to extend the class of maps which are particular
forms of the universal map we'll also consider $K$ not as a factor but 
as a parameter. Let's consider the following generating equation:
\begin{equation}
\dot{x}+G_K(x(t- \Delta T)) \sum^{\infty}_{n=-\infty} \delta \Bigl(\frac{t}{T}-(n+\varepsilon)
\Bigr)=0,
\label{UM1Ddif}
\end{equation}
where $0 < \varepsilon<1$ and  $0< \Delta <1$
with the initial condition:
\begin{equation}
x(0)=x_0.
\label{UM1DIC}
\end{equation}
1D analog of Eq. (\ref{Volt2Dxp}) (for
$(n+\varepsilon)T<t<(n+1+\varepsilon)T$) 
can be written as
\begin{equation}
x(t)=x_0  -  
T\sum^{n}_{k=0} G_K(x[ T(k+\varepsilon-\Delta)]).
\label{Volt1Dx}
\end{equation}
From the fact that $\dot{x}=0$ for $t \in
(T(k+\varepsilon-1), T(k+\varepsilon))$ it follows that 
$x[ T(k+\varepsilon-\Delta)]= x(Tk) $
and the corresponding 1D map can be written as a map  with full memory 
\begin{equation}
x_{n+1}= x_0 -  
T\sum^{n}_{k=0} G_K(x_k).
\label{UM1Dfmem}
\end{equation}
From Sec.~\ref{sec:1.2} it follows that this map can be written 
as the 1D form of
the universal map with one-step memory
\begin{equation}
x_{n+1}= x_n - T G_K(x_n).
\label{UM1D}
\end{equation}
It would be impossible to derive the logistic map from Eq.~(\ref{UM1D})  if
$K$ were a factor, but from the present form the logistic map can  
be obtained by assuming
\begin{equation}
G_K(x)=G_{lK}(x)=\frac{1}{T}[x-Kx(1-x)].
\label{LMG}
\end{equation}

In \cite{PopBioBook2001,PB3} Eq.~(\ref{UM1Ddif})  with no time delay, no delta
functions, and $G_K(x)$ defined by  Eq.~(\ref{LMG}) is used as 
one of the most general models in population biology and epidemiology.
Three terms in $G_K(x)$ represent a growth rate proportional 
to the current population,
restrictions due to the limited resources, and the death rate. 
The logistic map appears and plays an important role not only in
population biology but also in economics, condensed matter physics, and
other areas of science  \cite{PB1,PB3}. 
In population biology and epidemiology time delays can be related 
to the time of the development of an infection in a body until a person 
becomes infectious, or to the time of the development of an embryo.
For the importance of time delay in many scientific applications of the 
logistic map see e.g. Ch.~3 from  \cite{PB3} and Ch. 3 from \cite{PB1}. 
Changes which occur as periodically following discrete events
can be modeled by the delta function. 

The n-dimensional universal map can be derived from the following
generating equation:
\begin{equation}
\frac{d^nx}{dt^n}+G_K(x(t- \Delta)) \sum^{\infty}_{k=-\infty} \delta
\Bigl(t-(k+\varepsilon)
\Bigr)=0,   
\label{UM1D2Ddif}
\end{equation}
where $n \ge 0$, $n\in\mathbb{Z}$, and   $ \varepsilon > \Delta > 0$
in the limit $\varepsilon  \rightarrow 0$. This means that in
the general case time delay is not essential. 
Without losing the generality,  in Eq.~(\ref{UM1D2Ddif}) we assumed  $T=1$.
The case $T \ne 1$ is considered in \cite{DNC} and 
can be reduced to this case by rescaling the time variable and the map
generating function $G_K(x)$.
In Sec.~\ref{AFM}    $T$ denotes periods of trajectories. 
The 2D universal map Eqs.~(\ref{UMp})~and~(\ref{UMx}) 
corresponds to $n=2$ and the 1D universal map (\ref{UM1D})
corresponds to $n=1$. In the consistent introduction of fractional
derivatives integer derivatives appear as the limits when the order of 
a fractional derivative assumes an integer value. Correspondingly, the general
form of the n-dimensional universal map appears if we assume an integer
value of $\alpha$ in the general form of the fractional universal map.
In the following sections we'll consider the general forms of the
fractional universal map which will be derived from Eq.~(\ref{UM1D2Ddif})
with integer $n$ replaced by $\alpha \in \mathbb{R}$ ($\alpha \ge 0$).
The Riemann-Liouville universal map will be derived in Sec.~\ref{RLUM} and 
the Caputo  universal map will be derived in Sec.~\ref{CUM}.

\subsection{Riemann-Liouville Universal Map}
\label{RLUM}

The generating fractional differential equation for the 
Riemann-Liouville universal map can be written as 
\begin{equation}
_0D^{\alpha}_tx(t) +G_K(x(t- \Delta )) \sum^{\infty}_{n=-\infty} \delta \Bigl(t-(n+\varepsilon)
\Bigr)=0,  
\label{UM1D2DdifRL}
\end{equation}
 where $\varepsilon > \Delta > 0$, $\varepsilon  \rightarrow 0$, $0 \le
 N-1 < \alpha \le N$, $\alpha \in \mathbb{R}$, $N \in \mathbb{Z}$, 
and the initial conditions  
\begin{equation}
(_0D^{\alpha-k}_tx)(0+)=c_k,  \    \ k=1,...,N.
\label{UM1D2DdifRLic}
\end{equation}
The left-sided Riemann-Liouville fractional  derivative $_0D^{\alpha}_t
x(t)$ is defined for
$t>0$ \cite{KST,Podlubny,SKM} as 
\begin{equation}
_0D^{\alpha}_t x(t)=D^n_t \ _0I^{n-\alpha}_t x(t)=
\frac{1}{\Gamma(n-\alpha)} \frac{d^n}{dt^n} \int^{t}_0 
\frac{x(\tau) d \tau}{(t-\tau)^{\alpha-n+1}},
\label{RL}
\end{equation}
where $n-1 \le \alpha < n$, $D^n_t=d^n/dt^n$, and $ _0I^{\alpha}_t$ is a fractional integral.

For a wide class of functions $G_K(x)$  Eq.~(\ref{UM1D2DdifRL})  is equivalent
to the Volterra integral equation of the second
kind ($t>0$) (see \cite{KBT1,KBT2,KST,TarBook})
\begin{eqnarray}
&&x(t)= \sum^{N}_{k=1}\frac{c_k}{\Gamma(\alpha-k+1)}t^{\alpha -k}  \nonumber\\   
&&-\frac{1}{\Gamma(\alpha)} \int^{t}_0 d \tau \frac{G_K(x( \tau - \Delta ))}{( t-\tau )^{1-\alpha}} \sum^{\infty}_{k=-\infty}
\delta \Bigl(\tau-(k+\varepsilon)\Bigr).
\label{VoltRL}
\end{eqnarray}
Due to the presence of the delta function
the integral on the right side of Eq.~(\ref{VoltRL})
can be easily calculated \cite{DNC,MM9,MM8,TarBook} for $t>0$:
\begin{eqnarray}
&&x(t)= \sum^{N-1}_{k=1}\frac{c_k}{\Gamma(\alpha-k+1)}t^{\alpha -k} \nonumber\\   
&&-\frac{1}{\Gamma(\alpha)}  
\sum^{[t-\varepsilon]}_{k=0} \frac{G_K(x(k+\varepsilon-\Delta))}{( t-(k+\varepsilon))^{1-\alpha}} 
\Theta(t-(k+\varepsilon)),
\label{VoltRLeq}
\end{eqnarray}
where $\Theta(t)$ is the Heaviside step function.  In Eq.~(\ref{VoltRLeq})
we took into account that boundedness of $x(t)$ at $t=0$ requires $c_N=0$ and
$x(0)=0$. 
After the introduction  (see \cite{MM11})
\begin{equation}
p(t)= {_0D^{\alpha-N+1}_t}x(t)
\end{equation}
and
\begin{equation}
p^{(s)}(t)= {D^{s}_t}p(t), 
\end{equation}
where $s=0,1,...,N-2$,
Eq.~(\ref{VoltRLeq}) leads to
\begin{eqnarray}
&&p^{(s)}(t)= \sum^{N-s-1}_{k=1}\frac{c_k}{(N-s-1-k)!}t^{N -s-1-k} \nonumber \\
&&\hspace{-0.3cm}-\frac{1}{(N-s-2)!}  
\sum^{[t-\varepsilon]}_{k=0} G_K(x(k+\varepsilon-\Delta))( t-k )^{N-s-2}, 
\label{VoltRLeqp}
\end{eqnarray}
where $s=0,1,...,N-2$.
Assuming $x_n=x(n)$, for  $\varepsilon > \Delta > 0$  
Eqs.~(\ref{VoltRLeq})~and~(\ref{VoltRLeqp}) in the limit  $\varepsilon
\rightarrow 0$ give the equations of the Riemann-Liouville universal map

\begin{eqnarray}
&&x_{n+1}=  \sum^{N-1}_{k=1}\frac{c_k}{\Gamma(\alpha-k+1)}(n+1)^{\alpha -k} \nonumber \\  
&&-\frac{1}{\Gamma(\alpha)}\sum^{n}_{k=0} G_K(x_k) (n-k+1)^{\alpha-1}, 
\label{FrRLMapx} \\
&&p^s_{n+1}= \sum^{N-s-1}_{k=1}\frac{c_k}{(N-s-1-k)!} (n+1)^{N-s -1-k}
\nonumber \\  
&&-\frac{1}{(N-s-2)!}\sum^{n}_{k=0} G_K(x_k) (n-k+1)^{N-s-2}.
\label{FrRLMapp} 
\end{eqnarray}

\subsection{Caputo Universal Map}
\label{CUM}

Similar to (\ref{UM1D2DdifRL}), the generating fractional differential 
equation for the Caputo universal map can be written as 
\begin{equation}
_0^CD^{\alpha}_tx(t) +G_K(x(t- \Delta )) \sum^{\infty}_{n=-\infty} \delta \Bigl(t-(n+\varepsilon)
\Bigr)=0,    
\label{UM1D2DdifC}
\end{equation}
 where $\varepsilon > \Delta > 0$, $\varepsilon  \rightarrow 0$, $0 \le N-1
 < \alpha \le N$,  $\alpha \in \mathbb{R}$, $N \in \mathbb{Z}$, 
and the initial conditions 
\begin{equation}
(D^{k}_tx)(0+)=b_k,  \    \ k=0,...,N-1.
\label{UM1D2DdifCic}
\end{equation}
The left-sided Caputo fractional  derivative $_0^CD^{\alpha}_t x(t)$ is defined for
$t>0$ \cite{KST,Podlubny,SKM} as
\begin{equation}
_0^CD^{\alpha}_t x(t)=_0I^{n-\alpha}_t \ D^n_t x(t) =
\frac{1}{\Gamma(n-\alpha)}  \int^{t}_0 
\frac{ D^n_{\tau}x(\tau) d \tau}{(t-\tau)^{\alpha-n+1}},
\label{Cap}
\end{equation}
where $n-1 <\alpha \le n$.

For a wide class of functions $G_K(x)$  Eq.~(\ref{UM1D2DdifC})  is equivalent
to the Volterra integral equation of the second
kind ($t>0$) (see \cite{KBT1,KBT2,KST,TarBook})
\begin{equation}
x(t)= \sum^{N-1}_{k=0}\frac{b_k}{k!}t^{k} 
-\frac{1}{\Gamma(\alpha)} \int^{t}_0 d \tau \frac{G_K(x( \tau - \Delta ))}{( t-\tau )^{1-\alpha}} \sum^{\infty}_{k=-\infty}
\delta \Bigl(\tau-(k+\varepsilon)\Bigr).
\label{VoltC}
\end{equation}
Integration of this equation gives for $t>0$
\begin{equation}
x(t)=  \sum^{N-1}_{k=0}\frac{b_k}{k!}t^{k} 
-\frac{1}{\Gamma(\alpha)}  
\sum^{[t-\varepsilon]}_{k=0} \frac{G_K(x( k+\varepsilon-\Delta  ))}{( t-(k+\varepsilon) )^{1-\alpha}} 
\Theta(t-(k+\varepsilon)).
\label{VoltCeq}
\end{equation}
After the introduction $x^{(s)}(t)=D^s_tx(t)$ the Caputo  universal map
can be derived in the form (see \cite{TarBook})
{\setlength\arraycolsep{0.5pt}
\begin{equation}
x^{(s)}_{n+1}= \sum^{N-s-1}_{k=0}\frac{x^{(k+s)}_0}{k!}(n+1)^{k} 
-\frac{1}{\Gamma(\alpha-s)}\sum^{n}_{k=0} G_K(x_k) (n-k+1)^{\alpha-s-1},
\label{FrCMapx}
\end{equation}
}
where $s=0,1,...,N-1$.

\section{$\alpha$-Families of Maps}
\label{AFM}

We'll  call  Eqs. (\ref{UM1D2DdifRL}) and (\ref{UM1D2DdifRLic})
with various map generating functions $G_K(x)$ the 
Riemann-Liouville universal map generating equations
and Eqs.~(\ref{FrRLMapx}) and (\ref{FrRLMapp}) the Riemann-Liouville 
$\alpha$-families of maps corresponding to the functions $G_K(x)$.
We'll  call  Eqs. (\ref{UM1D2DdifC}) and (\ref{UM1D2DdifCic})
with various map generating functions $G_K(x)$ the 
Caputo universal map generating equations
and Eqs.~(\ref{FrCMapx})  the Caputo 
$\alpha$-families of maps corresponding to the functions $G_K(x)$.

Fractional maps
Eqs.~\eqref{FrRLMapx},~\eqref{FrRLMapp},~and~\eqref{FrCMapx}
are maps with memory in which the next values of  map variables depend
on all previous values. An increase in  $\alpha$ corresponds to 
the increase in a map dimension.  It also corresponds to 
the increased power in the power-law dependence of  weights of  
previous states which imply  increased memory effects.
For $\alpha=1$ and $\alpha=2$ the corresponding maps are given by 
Eqs.~(\ref{UM1D}),~(\ref{UMp}),~and~(\ref{UMx}) with $T=1$ and $G_K(x)$
instead of $G(x)$. 
Eqs.~\eqref{FrRLMapx},~\eqref{FrRLMapp},~and~\eqref{FrCMapx} with
$\alpha=3$ and variables  $y=p$ and $z=\dot{p}$ produce the full-memory 3D Universal Map
\begin{eqnarray}
&&x_{n+1}=  \frac{z_0}{2}(n+1)^2 + y_0(n+1)+x_0
-\frac{1}{2}\sum^{n}_{k=0} G_K(x_k) (n-k+1)^{2}, \nonumber \\
&&y_{n+1}=  z_0(n+1) + y_0 -
\sum^{n}_{k=0} G_K(x_k) (n-k+1), \label{3DUMf}  \\
&&z_{n+1}=   z_0 -
\sum^{n}_{k=0} G_K(x_k), \nonumber
\end{eqnarray} 
which is equivalent to the one-step memory 
(Sec.~\ref{sec:1.2}) 3D universal map 
\begin{eqnarray}
&&x_{n+1}=  x_n-\frac{1}{2}G_K(x_n)+y_n+\frac{1}{2}z_n,  \nonumber \\
&&y_{n+1}=-G_K(x_n)+y_n+z_n,  \label{3DUM} \\
&&z_{n+1}=-G_K(x_n)+z_n, \nonumber 
\end{eqnarray} 
or
\begin{eqnarray}
&&x_{n+1}=  x_n+y_{n+1}-\frac{1}{2}z_{n+1},   \nonumber \\
&&y_{n+1}=y_n+z_{n+1}, \label{3DUMn} \\
&&z_{n+1}=-G_K(x_n)+z_n, \nonumber 
\end{eqnarray} 
which is a volume preserving map.
This map has fixed points  $z_0=y_0=G_K(x_0)=0$ and
stability of these points can be analyzed by considering the eigenvalues 
$\lambda$ of the
matrix (corresponding to the tangent map)
\[ \left( \begin{array}{ccc}
1-0.5\dot{G_K}(x_0) & 1 & 0.5 \\
-\dot{G_K}(x_0) & 1 & 1 \\
-\dot{G_K}(x_0) & 0 & 1 \end{array} \right).\]
The only case in which the fixed points could be stable is  $\dot{G_K}(x_0)=0$,
when $\lambda_1=\lambda_2=\lambda_3=1$. From Eq.~(\ref{3DUMn}) it follows that 
the only $T=2$ points are the fixed points.

The investigation of the integer members of the $\alpha$-families of maps 
is a subject of ongoing research.
From the examples of maps with the values of $\alpha$ equal to one, 
two, and three we see that integer values of $\alpha$ correspond 
to the degenerate cases in which map equations can be written
as maps with full memory. They are equivalent to $\alpha$-dimensional  
one-step memory maps
in which map variables at each step accumulate 
information about all previous states of the corresponding systems.

Corresponding to the fact that in the $\alpha=2$ case 
the 2D universal family of maps  produces the 
standard map if $G_K(x)=K\sin(x)$ (see Eqs.~(\ref{SM})) and
in the $\alpha=1$ case the logistic map results from $G_K(x)=x-Kx(1-x)$
(see Eqs.~(\ref{UM1D})~and~(\ref{LMG})), we'll call:
\begin{itemize}
\item{the Riemann-Liouville 
$\alpha$-family of maps  Eqs.~(\ref{FrRLMapx}) and (\ref{FrRLMapp}) with  
$G_K(x)=K\sin(x)$ the \bf{standard $\alpha$-RL-family of maps};}
\item{the Caputo 
$\alpha$-family of maps  Eqs.~(\ref{FrCMapx}) with  
$G_K(x)=K\sin(x)$ the \bf{standard $\alpha$-Caputo-family of maps};}
\item{the Riemann-Liouville 
$\alpha$-family of maps with $G_K(x)=x-Kx(1-x)$    
the \bf{logistic $\alpha$-RL-family of maps};}
\item{the Caputo 
$\alpha$-family of maps  with  
 $G_K(x)=x-Kx(1-x)$  the \bf{logistic $\alpha$-Caputo-family of maps}}.
\end{itemize}

For $\alpha=0$ the solution of Eq.~\eqref{UM1D2Ddif}
and correspondingly, the universal map is identically zero.
For $\alpha <1 $ the  Riemann-Liouville 
$\alpha$-families of maps  Eqs.~\eqref{FrRLMapx} and (\ref{FrRLMapp}) 
corresponding to the functions $G_K(x)$  satisfying 
the condition $G_K(0)=0$, which 
is true for the standard and logistic $\alpha$-RL-families of maps,
also produces identically zero.

\subsection{Integer-Dimensional Standard and Logistic Maps}
\label{IntStLog}

In general, properties of fractional maps converge to 
the corresponding properties of integer maps when $\alpha$ 
approaches integer values.
To better understand properties of fractional maps  
we'll start with the consideration of the integer members of the
corresponding families of maps.

\subsubsection{One-Dimensional Logistic and Standard  Maps}
\label{1DStLog}

\begin{figure}[!t]
\includegraphics[width=1.0\textwidth]{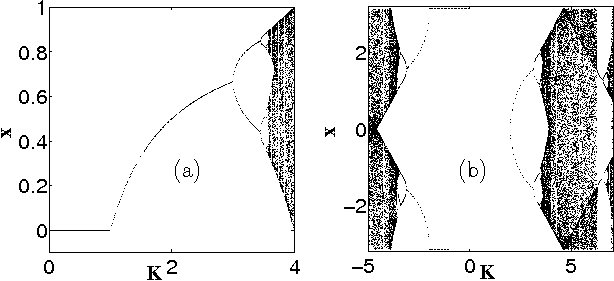}
\vspace{-0.25cm}
\caption{(a) The bifurcation diagram for the  logistic map 
$x=Kx(1-x)$. 
(b) The bifurcation diagram for the 1D standard map (circle map) Eq.~(\ref{SM1D}). 
}
\label{BD1D}
\end{figure}

The one-dimensional logistic map Eq.~(\ref{LM}) is one of the best
investigated maps.  
This map has been used as a playground for
investigation of the essential property of nonlinear systems - 
transition from order to chaos through a sequence of period-doubling
bifurcations, which is called cascade of bifurcations, and  
scaling properties of the corresponding systems (see
\cite{LM1,LM2,LM3,LM4,LM6}). In our investigation of fractional maps we'll
use the well known stability properties of the logistic map  
(see \cite{LM5}), which for $0<K<4$ are summarized in the bifurcation
diagram in Fig.~\ref{BD1D}(a).
The $x=0$ fixed point (sink) is stable for  $K < 1$, 
the $(K-1)/K$ fixed point (sink) is stable for $1 < K < 3$, the  $T=2$ 
sink is stable
for $3 \le K < 1-\sqrt{6} \approx 3.449$, the $T=4$ sink is stable
when $3.449 < K < 3.544$, and the onset of
chaos as a result of the period-doubling cascade 
of bifurcations occurs at $K  \approx 3.56995$.

\begin{figure}
\centering
\includegraphics[width=1.0\textwidth]{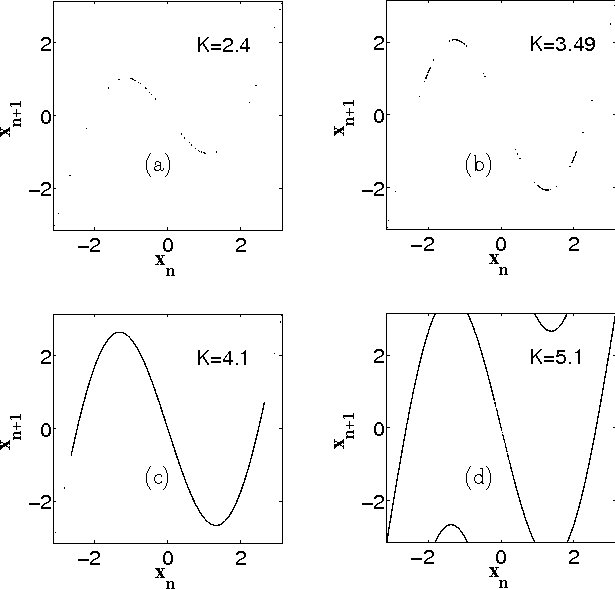}
\caption{\label{SM1Dpp} Attractors in the one-dimensional standard map;
$x_n$ vs. $x_{n+1}$ plots (seven trajectories with different initial
conditions in each plot):
(a). $K=2.4$; antisymmetric $T=2$ sink.
(b). $K=3.49$; $T=4$ trajectories.
(c). $K=4.1$; proper attractor (width of the chaotic area is less than $2\pi$).
(d). $K=5.1$; improper attractor (width of the chaotic area is $2\pi$).
}
\end{figure}

The one-dimensional standard map ($\alpha=1$) considered on a circle
\begin{equation}
x_{n+1}= x_n - K \sin (x_n), \ \ \ \ ({\rm mod} \ 2\pi ) 
\label{SM1D} 
\end{equation}
is a particular form of the circle map with zero driving phase.
It has  attracting fixed points $2\pi  n$  for $0<K \le K_{c1}(1)=2$ and 
$ \pi + 2 \pi n$ when $-2 \le K < 0$ (for the
bifurcation diagram of the 1D standard map see Fig.~\ref{BD1D}(b)). 
The antisymmetric $T=2$ sink 
\begin{equation}
x_{n+1}= -x_n  
\label{T21DSym} 
\end{equation}
is stable for $2 < |K| < \pi$, while
$x_{n+1} = x_n+\pi$ 
two sinks ($T=2$) are
stable when $\pi < |K| < \sqrt{\pi^2+2} \approx 3.445$. 
The stable $T=4$ sinks appear at $|K| \approx 3.445$ and
the sequence of bifurcations $T=4$ $\rightarrow$ $T=8$ at $K \approx 3.513$,
$T=8$ $\rightarrow$ $T=16$ at $K \approx 3.526$, 
and so on leads to the transition to chaos at
$K \approx 3.532$. 
Antisymmetric $T=2$ trajectories ($K=2.4$),  $T=4$ trajectories ($K=3.49$), and 
two cases of chaotic trajectories ($K=4.1$ and $K=5.1$) 
are presented in Fig.~\ref{SM1Dpp}. 
In the 1D standard  map with $K>0$ the full phase space $x \in [-\pi, \pi]$ 
becomes involved in chaotic
motion (we'll call this case ``improper attractor'') when the maximum of 
the function $f_K(x)=x-K \sin x$ is equal to $\pi$ which
occurs at $K_{max1D}= 4.603339$ when $x_{max1D}=-1.351817$
(see Figs.~\ref{SM1Dpp}~(c)~and~(d)).
Narrow bands with $|K|$ above $2\pi |n| $ (see   Fig.~\ref{BD1D}(b) for
$K>2\pi$) are accelerator mode bands with zero acceleration within which
in the unbounded space (no ${\rm mod} \ 2\pi $) $x$ is
increasing/decreasing with the rate equal approximately to $ 2\pi |n|$.

\subsubsection{Two-Dimensional Logistic and Standard  Maps}
\label{2DStLog}

The two-dimensional  logistic map 
\begin{eqnarray}
&& p_{n+1}= p_n+Kx_n(1-x_n)-x_n, 
\nonumber \\
&& x_{n+1}= x_n + p_{n+1}
\label{LFMalp2}
\end{eqnarray}
is  a  quadratic area preserving map. Its phase space
contains  stable elliptic islands and chaotic areas (no attractors).  
Quadratic area preserving maps which have a stable fixed point at the origin  
were investigated by H\'enon \cite{Henon69}  (for a recent review
on 2D quadratic maps see \cite{ZeraS2010}). To investigate the logistic
$\alpha$-families of maps we need to know the evolution of the periodic
points of the 2D logistic map with the increase of the map parameter $K$.  
For $K \in (-3,1)$ the map Eq.~\eqref{LFMalp2} has the stable fixed point
$(0,0)$  which turns into the  fixed point $((K-1)/K,0)$ stable 
for  $K \in (1,5)$.
The $T=2$ elliptic point
{\setlength\arraycolsep{0.5pt}
\begin{eqnarray}
&&x = \frac{K+3 \pm \sqrt{(K+3)(K-5)}}{2K},  \nonumber \\ 
&&p=\pm \frac{\sqrt{(K+3)(K-5)}}{K}
\label{LFMalp2T2}
\end{eqnarray} 
}
\begin{figure}[!t]
\includegraphics[width=1.0\textwidth]{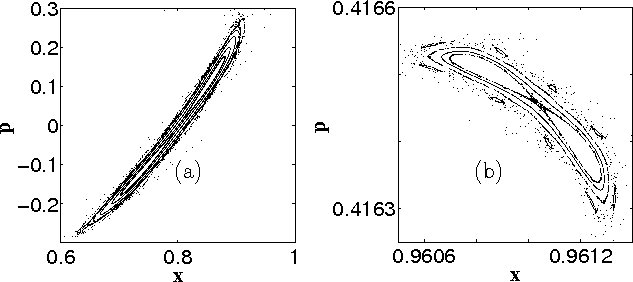}
\vspace{-0.25cm}
\caption{ 
Bifurcations in the 2D Logistic Map:
(a) $T=1$ $\rightarrow$ $T=2$ bifurcation at $K = 5$ ($K=5.05$ 
on the figure). 
(b) $T=8$ $\rightarrow$ $T=16$ bifurcation at $K \approx 5.5319$  ($K=5.53194$ 
on the figure).   
}
\label{FigLog2D}
\end{figure}
is stable for $-2 \sqrt{5}+1<K<-3$ and $5<K<2 \sqrt{5}+1$.
The period doubling cascade of bifurcations (for $K>0$) 
follows the scenario of 
the elliptic-hyperbolic point transitions with the births
of the double periodicity islands inside the original island 
which has been investigated in \cite{Schmidt} and applied to investigate the 
standard map stochasticity at low values of the map parameter.
Further bifurcations  in the 2D logistic map, 
$T=2$ $\rightarrow$ $T=4$ at $K \approx 5.472$,
$T=4$ $\rightarrow$ $T=8$ at $K \approx 5.527$, $T=8$ $\rightarrow$ $T=16$
at $K \approx 5.5319$, $T=16$ $\rightarrow$ $T=32$ at $K \approx 5.53253$,
etc., and the corresponding decrease in the areas of the islands of stability
(see Fig.~\ref{FigLog2D}) lead to chaos.

The  two-dimensional standard map on a torus  
Eq.~(\ref{SM}) (Chirikov standard map) 
is one of the best investigated 2D maps. It demonstrates 
a universal generic behavior of the area-preserving maps whose
phase space is divided into elliptic islands of stability and areas of 
chaotic motion (see, e.g., \cite{Chir,LL}). 
Elliptic islands of the standard map in the case of  
the standard $\alpha$-families of maps with $1 < \alpha < 2$ 
evolve into periodic sinks (see Sec.~\ref{BN12}).
Properties of phase space and  appearance 
of different types of attractors in the fractional case, as in the case
of the fractional logistic map, 
are connected to the evolution (with 
the increase in parameter $K$) of the 2D standard map's islands 
originating from the stable (for $K<4$) fixed point (0,0). 
At $K=4$ the fixed point becomes unstable (elliptic-hyperbolic point transition  
\cite{Schmidt}) 
and two elliptic islands around the stable for $4 < K <2 \pi$  
period 2 antisymmetric 
point 
\begin{equation}\label{SimP2D}
p_{n+1}=-p_n, \    \ x_{n+1}=-x_n
\end{equation}
appear. At $K=2 \pi$ this point transforms into two  $T=2$ points 
\begin{equation}
p_{n+1}=-p_n, \    \  x_{n+1}=x_n-\pi, 
\label{ASimP2D}
\end{equation}
which are stable when $2 \pi <K<6.59$.
These points transform into $T=4$ stable elliptic points  
at  $K \approx 6.59$ and the period
doubling cascade of bifurcations leads to the disappearance of
islands of stability in the chaotic sea at $K \approx 6.6344$
\cite{Chir,LL}. 
The 2D standard map has a set of bands for $K$ above $2\pi n$ 
of the accelerator mode sticky islands in which the momentum $p$ increases
proportionally to the number of iterations $k$ and the coordinate $x$ 
increases as $k^2$. 
The role of accelerator mode islands (for $K$ above $2\pi$) in
the anomalous diffusion and the corresponding fractional kinetics is well
investigated (see, for example, \cite{ZasBook,ZE1997}).

\subsubsection{Three-Dimensional Logistic and Standard  Maps}
\label{3DStLog}

Eq.~(\ref{3DUMn}) with $G_K(x)=x-Kx(1-x)$ (see Eq.~(\ref{LMG}) produces
the 3D logistic map
\begin{eqnarray}
&&x_{n+1}=  x_n+y_{n+1}-\frac{1}{2}z_{n+1},  \nonumber \\
&&y_{n+1}=y_n+z_{n+1}, \label{3DLMn} \\
&&z_{n+1}=Kx_n(1-x_n)-x_n+z_n. \nonumber 
\end{eqnarray} 
Three-dimensional quadratic volume preserving  maps were investigated in 
\cite{Moser1994,LoMeiss1998}. Everything stated in Sec.~\ref{AFM}
for the 3D universal map is still valid for the 3D logistic map.

The three-dimensional standard map with $G_K(x)=K \sin(x)$
\begin{eqnarray}
&&x_{n+1}=  x_n+y_{n+1}-\frac{1}{2}z_{n+1},\   \ ({\rm mod} \ 2\pi ), 
\nonumber \\
&&y_{n+1}=y_n+z_{n+1}, \   \  ({\rm mod} \ 2\pi ),       \label{3DSMn} \\
&&z_{n+1}=-K\sin(x_n)+z_n, ,\   \ ({\rm mod} \ 4\pi ) \nonumber 
\end{eqnarray}  
has unstable fixed points $(2\pi n, 2\pi m, 4 \pi k)$ and $(2\pi n+ \pi , 2\pi m, 4 \pi k)$ , 
$n \in  \mathbb{Z}$, $m \in  \mathbb{Z}$, $k \in  \mathbb{Z}$. 
Ballistic points $K\sin(x)=-4\pi n$,
$y=2\pi m$, $z=4\pi k$, which appear for $|K| \ge 4\pi$, are also unstable.

Stability of $T=2$ ballistic points is defined by the eigenvalues of the matrix 
\[ \left( \begin{array}{ccc}
1-0.5K \cos x_1 & 1 & 0.5 \\
-K \cos x_1 & 1 & 1 \\
-K \cos x_1 & 0 & 1 \end{array} \right) \times
\left( \begin{array}{ccc}
1-0.5 K \cos x_2 & 1 & 0.5 \\
-K \cos x_2 & 1 & 1 \\
-K \cos x_2 & 0 & 1 \end{array} \right).\]
For the period two on the torus ballistic points   
\begin{eqnarray}
&&z_1,\  \ y_1=\frac{z_1}{2}-\pi(2n+1),\  \ K \sin x_1=2z_1, \nonumber \\  
&& z_2=-z_1, \  \ y_2=-\frac{z_1}{2}-\pi(2n+1),\  \ x_2=x_1-\pi(2n-1),  
\label{3DT2StablePoint}
\end{eqnarray}  
where $n \in  \mathbb{Z}$, the  eigenvalues are
\begin{equation}
 \Bigl\{1, \  \  \frac{1}{8}(8-K^2 \cos^2 x_1 \pm K \cos x_1 \sqrt{K^2\cos^2
  x_1-16})  \Bigr\}
\label{3DT2StablePointEigenV}
\end{equation} 
\begin{figure}
\centering
\includegraphics[width=1.0\textwidth]{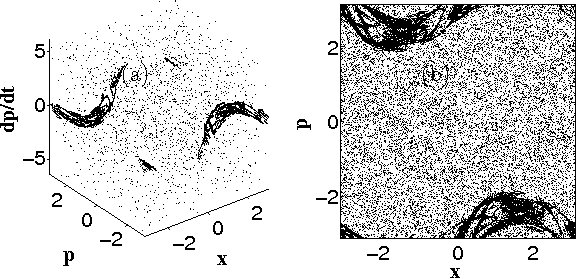}
\caption{\label{3D} Phase space of the 3D standard map (\ref{3DSMn}) with $K=3$:
(a). Three dimensional phase space. (b). A projection of the 3D phase space
on the $x$-$y$ plane.   
}
\end{figure} 
Ballistic $T=2$ points are stable along a line defined by
Eqs.~(\ref{3DT2StablePoint}) for all values of $z$ satisfying the condition
\begin{equation}
K^2-16<4z^2<K^2.
\label{3DT2zCond}
\end{equation}
An example of the phase space for $K=3$ in three dimensions and its projection
on the $x$-$y$ plane is given in Fig.~\ref{3D}. For this value of $K$
ballistic $T=2$ points are stable when $-1.5<z<1.5$
and the space around the line of stability presents a series of islands
(invariant curves), islands around islands, and separatrix layers. 
When $K \rightarrow 0$, the
volume of the regular motion shrinks. When $K$ is small, the line of the 
stable $T=2$ ballistic points exists for $-K/2<z<K/2$. 
A different form of the 3D volume preserving standard map was introduced
and investigated in detail in \cite{DM}.

\subsection{$\alpha$-Families of Maps ($0<\alpha<1$)}
\label{LT1}

As we mentioned at the end of Sec.~\ref{AFM}, members of the logistic and 
standard $\alpha$-families of maps corresponding to $\alpha=0$ and 
RL-families' members with  $0<\alpha<1$ are identically zeros.
The only fractional logistic and standard maps with $0<\alpha<1$ 
which are not identically zeros are $\alpha$-Caputo-families of maps.
The $\alpha$-Caputo-universal map ($0<\alpha<1$)
\begin{equation}     
x_{n+1}=  x_0- 
\frac{1}{\Gamma(\alpha)}\sum^{n}_{k=0} G_K(x_k) (n-k+1)^{\alpha-1}
\label{FrCMapxlt1}
\end{equation}
in  the limit  $\alpha \rightarrow 1$ is identical to the one-dimensional 
universal map Eq.~(\ref{UM1D}) and in this limit properties of 
fractional maps are similar to properties of the corresponding 1D maps.
Eq.~(\ref{FrCMapxlt1}) with  $G_K(x)=x-Kx(1-x)$ is the logistic 
$\alpha$-Caputo-family of maps for $0<\alpha<1$
\begin{equation}
x_{n}=  x_0+ 
\frac{1}{\Gamma(\alpha)}\sum^{n-1}_{k=0} \frac{Kx_k(1-x_k)-x_k}{(n-k)^{1-\alpha}}
\label{FrCMapLM}
\end{equation}
and with $G_K(x)=K\sin(x)$ is  the standard $\alpha$-Caputo-family of maps for $0<\alpha<1$
\begin{equation}
x_{n}=  x_0- 
\frac{K}{\Gamma(\alpha)}\sum^{n-1}_{k=0} \frac{\sin{x_k}}{(n-k)^{1-\alpha}},
 \   \  ({\rm mod} \ 2\pi )
\label{FrCMapSM}
\end{equation}
These maps are one-dimensional maps
with  power-law decreasing memory \cite{DNC}.
The bifurcation diagrams for these 
maps are similar to the corresponding diagrams for the $\alpha=1$ case
Fig.~\ref{BD1D}.
\begin{figure}[!t]
\begin{center}
\includegraphics[width=0.95\textwidth]{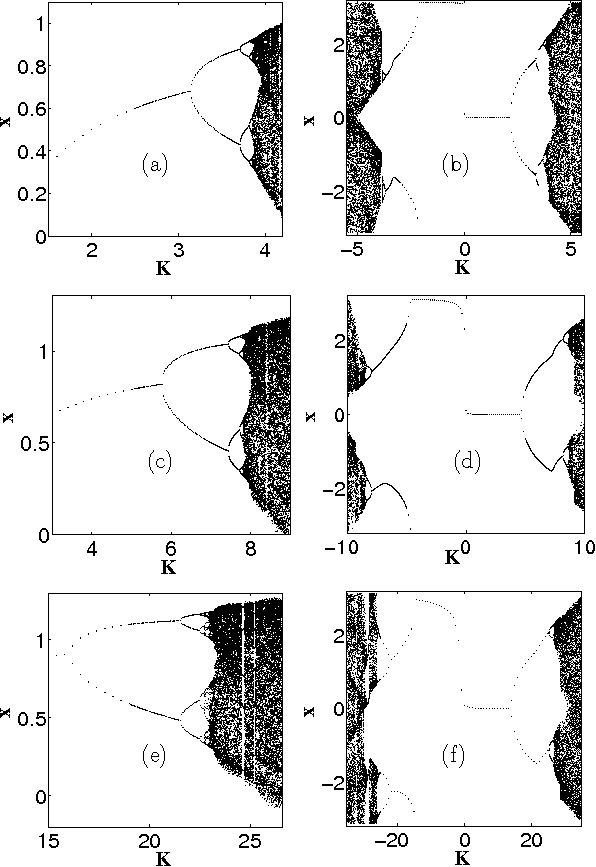}
\vspace{-0.25cm}
\caption{Bifurcation diagrams for  the logistic
and standard $\alpha$-Caputo-families of maps 
with $0<\alpha<1$. In (a)-(f) the bifurcation 
diagrams obtained after performing 
$10^4$ iterations on a single trajectory with $x_0=0.1$ 
for various values of $K$.  
(a), (c), and (e) - the logistic $\alpha$-Caputo-family. 
(b), (d), and (f) - the standard $\alpha$-Caputo-family. 
In (a) and (b) $\alpha=0.8$.
In (c) and (d) $\alpha=0.3$.
In (e) and (f) $\alpha=0.1$.
}
\label{LowAlpBif}
\end{center}
\end{figure}
A decrease in $\alpha$ and the corresponding decrease in weights of the
earlier states (decrease in memory effects) leads to the stretchiness 
of the corresponding bifurcation diagrams along the parameter $K$-axis 
and this stretchiness increases as $\alpha$ gets smaller Fig.~\ref{LowAlpBif}.

\begin{figure}[!t]
\includegraphics[width=0.98\textwidth]{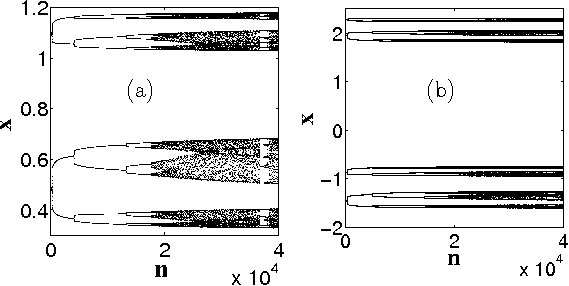}
\vspace{-0.25cm}
\caption{Cascade of bifurcations type trajectories in the logistic
and standard $\alpha$-Caputo-families of maps 
with $\alpha=0.1$. 
(a) The fractional logistic map with   $\alpha=0.1$ and $K=22.65$.
(b) The fractional standard map with   $\alpha=0.1$ and $K=26.65$.
}
\label{CBTT1D}
\end{figure}

Within a band of values of $K$, above the value which corresponds to the
appearance of  $T=4$ trajectories, map trajectories are attracting 
cascade of bifurcations type trajectories (CBTT) (see Fig.~\ref{CBTT1D}).
On CBTT an increase in the number of map iterations 
leads to the change in the map's stability properties.  A trajectory which
converges to a $T>4$ periodic point or becomes a chaotic trajectory 
(depending on the value of $K$) evolves according to a certain scenario:
it first converges to a $T=4$ point; then it bifurcates, always at the
same place for the given values of the parameter $K$ and the order
$\alpha$, and converges to a $T=8$ trajectory; 
then to a $T=16$ trajectory; and so on. 
Power-law decaying memory with power $\beta \approx 0.9$ corresponding to
small values of $\alpha \approx 0.1$ 
(see Sec.~\ref{sec:1.1}) appears in biological applications. Attracting CBTT 
in, for example, adaptive biological systems may represent 
not simply a change of a state of a biological system 
according to a change in a parameter, but rather a change in the evolution of 
the system according to the change in the parameter. Examples of 
CBTT in the logistic and standard $\alpha$-Caputo-families of maps 
with $\alpha=0.1$ are presented in  Fig.~\ref{CBTT1D}.

\begin{figure}[!t]
\includegraphics[width=1\textwidth]{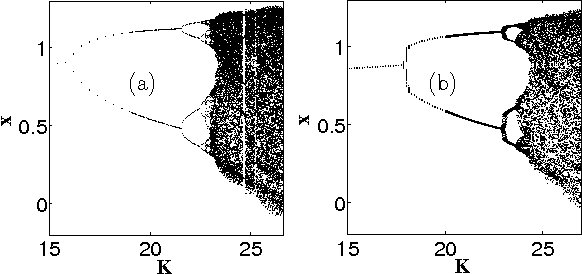}
\vspace{-0.25cm}
\caption{Dependence of bifurcation diagrams of the fractional maps 
on the number of iterations
on a single trajectory used in their calculation.  
Bifurcation diagrams for the fractional logistic map with $\alpha=0.1$.
(a) 10000 iterations on each trajectory. 
(b) 100 iterations on each trajectory.
}
\label{BifOfNI}
\end{figure}

It also should be noted that bifurcation diagrams of the fractional maps 
depend on the number of iterations used in their calculations.
This is a consequence of the existence of CBTT. Trajectories 
which after 100 iterations converged to a fixed point in 
Fig.~\ref{BifOfNI}(b) after 10000 iterations became $T=2$ trajectories
in Fig.~\ref{BifOfNI}(a). With an increase in the number of iterations
the whole bifurcation diagram shifts to the left.  

\subsection{$\alpha$-Families of Maps ($1<\alpha<2$)}
\label{BN12}

For $1<\alpha<2$ the logistic and standard $\alpha$-families of maps
assume the following forms:
\begin{itemize}
\item{
The RL-standard map on a cylinder 
\begin{eqnarray}
&&p_{n+1} = p_n - K \sin x_n ,  \label{FSMRLp}  \\
&&x_{n+1} = \frac{1}{\Gamma (\alpha )} 
\sum_{i=0}^{n} p_{i+1}V^1_{\alpha}(n-i+1) 
, \ \ \ \ ({\rm mod} \ 2\pi ), \label{FSMRLx} 
\end{eqnarray} 
where 
\begin{equation} \label{V1}
V^k_{\alpha}(m)=m^{\alpha -k}-(m-1)^{\alpha -k}. 
\end{equation}
This map requires the initial condition $x_0=0$ and
can't be considered on a torus.}
\item{The Caputo-standard map on a torus
\begin{eqnarray}
&& p_{n+1} = p_n 
-\frac{K}{\Gamma (\alpha -1 )} 
\Bigl[ \sum_{i=0}^{n-1} V^2_{\alpha}(n-i+1) \sin x_i 
+ \sin x_n \Bigr],\ \ ({\rm mod} \ 2\pi ),  \label{FSMCp} \\
&&x_{n+1} = x_n + p_0  
-\frac{K}{\Gamma (\alpha)} 
\sum_{i=0}^{n} V^1_{\alpha}(n-i+1) \sin x_i,\ \ ({\rm mod} \ 2\pi ). 
\label{FSMCx}
\end{eqnarray} 
} 
\item{
The RL-logistic map  
\begin{eqnarray}
&&p_{n+1} = p_n - Kx_n (1-x_n)-x_n,  \label{LMRLp}  \\
&&x_{n+1} = \frac{1}{\Gamma (\alpha )} 
\sum_{i=0}^{n} p_{i+1}V^1_{\alpha}(n-i+1), \label{LMRLx} 
\end{eqnarray} 
which requires the initial condition $x_0=0$.}
\item{
The Caputo-logistic map  
\begin{eqnarray}
&&x_{n+1}=x_0+ p(n+1)^{k} 
-\frac{1}{\Gamma(\alpha)}\sum^{n}_{k=0} [x_k-Kx_k(1-x_k)] (n-k+1)^{\alpha-1}, 
\label{LMCx}  \\
&&p_{n+1}=p_0 
-\frac{1}{\Gamma(\alpha-1)}\sum^{n}_{k=0} [x_k-Kx_k(1-x_k)] (n-k+1)^{\alpha-2}.
\label{LMCp}
\end{eqnarray} 
Here and in Eqs.~(\ref{FSMCp})~and~(\ref{FSMCx})   we assumed $x \equiv
x^0$ and  $p \equiv x^1$  in the Caputo universal map Eq.~(\ref{FrCMapx}).
}
\end{itemize}
The fractional standard maps  
Eqs.~(\ref{FSMRLp}),~(\ref{FSMRLx}),~(\ref{FSMCp}),~and~(\ref{FSMCx}) are
well investigated 
(see \cite{ME1,DNC,ME3,ME4}) 
and the logistic maps are the subject of ongoing research.

\begin{figure}[!t]
\begin{center}
\includegraphics[width=0.7\textwidth]{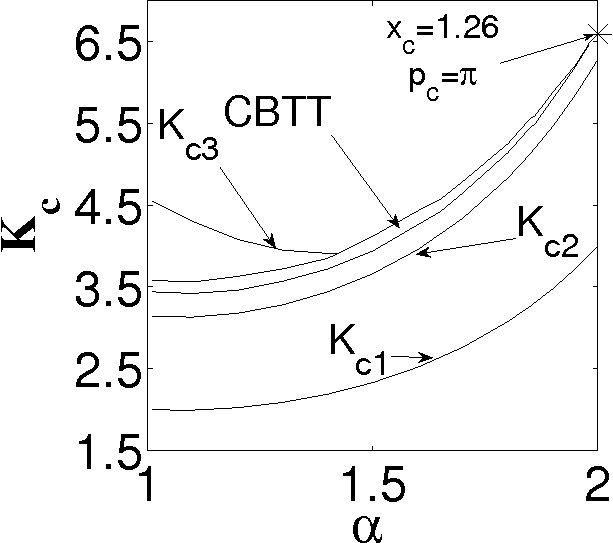}
\end{center}
\vspace{-0.25cm}
\caption{ Bifurcations in the standard $\alpha$-families of maps
with $1<\alpha<2$.  
Below $K=K_{c1}$ curve the fixed point $(0,0)$  is stable. It becomes 
unstable at  $K=K_{c1}$ and gives birth to the antisymmetric $T=2$
sink which is stable for $K_{c1}<K<K_{c2}$. A pair of the $T=2$ sinks with
$x_{n+1}=x_n-\pi$,  $p_{n+1}=-p_n$ is stable in the band above $K=K_{c2}$
curve. Cascade of bifurcations type trajectories (CBTT) 
appear and exist in the narrow band which ends 
at the cusp at the top right corner of the figure. 
$(x_c,p_c)$ is the point at which the standard
map's ($\alpha=2$) $T=2$ elliptic points with $x_{n+1}=x_n-\pi$,  
$p_{n+1}=-p_n$ become unstable and bifurcate. In the area below $K_{c3}$
(above the CBTT band) the chaotic attractor is restricted to a band whose
width is less than $2\pi$. On the upper curves and above them 
the full phase space is chaotic.}
\label{figBif}
\end{figure}

Evolution of trajectories in fractional maps depends on two parameters:
the map parameter $K$ and the fractional order $\alpha$. Fig.~\ref{figBif}
reflects this dependence in the case of the  standard $\alpha$-families of maps
with $1<\alpha<2$. 

\subsubsection{ $T=2$ Antisymmetric Sink}
\label{T2}

It is obvious that the fractional standard and logistic maps 
have the fixed points at the origin $(0,0)$.
But we'll start the fractional maps' phase space analysis   
with the consideration 
of the $T=2$ antisymmetric sinks. We'll present most of the analysis for
the fractional RL-standard map (Fig.~\ref{figBif}). Results of numerical
simulations suggest that the fractional Caputo-standard map has similar
properties and the results for the logistic map are submitted for 
publication.

\begin{figure}
\centering
\includegraphics[width=0.95\textwidth]{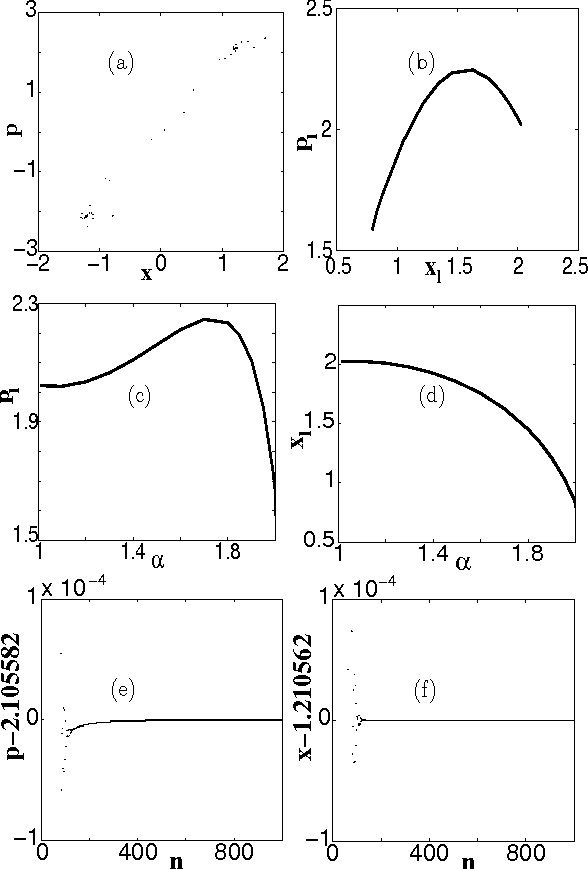}
\caption{\label{T2sink}  The RL-standard map's period 2 sink: 
(a). An example of the $T=2$ attractor for $K=4.5$, $\alpha =1.9$.
One trajectory with $x_0=0$,  $p_0=0.513$.
(b). $p_l$ of $x_l$ for the case of $K=4.5$. 
(c).  $p_l$ of $\alpha $ for the case of $K=4.5$.
(d). $x_l$ of $\alpha $ for the case of $K=4.5$.
(e). $p_n-p_l$ for the trajectory in (a). After 1000 iterations 
$|p_n-p_l| < 10^{-7}$.
(f). $x_n-x_l$ for the trajectory in (a). After 1000 iterations 
$|x_n-x_l| < 10^{-7}$.
 }
\end{figure}

The 1D standard map has the $T=2$ antisymmetric sink Eq.~(\ref{T21DSym})
and the 2D standard map has the $T=2$ antisymmetric 
elliptic point Eq.~(\ref{SimP2D}). Numerical experiments
(Fig.~\ref{T2sink}) show that the antisymmetric $T=2$ sinks  persist in the
fractional standard maps with $1<\alpha<2$.
In the RL-standard map these
sinks in the RL-standard map attract most of the trajectories 
with small $p_0$. 
Assuming the existence of an antisymmetric $T=2$ sink
\begin{equation} \label{RLT2sink} 
p_n=p_l(-1)^n, \  \ x_n=x_l(-1)^n ,
\end{equation} 
it is possible to calculate the coordinates of its attracting 
points $(x_l, p_l)$ 
and $(-x_l, -p_l)$.
In the limit $n \rightarrow \infty$  
Eqs.~(\ref{FSMRLp})~and~(\ref{FSMRLx}) can be written as
\begin{eqnarray} 
&&p_l = \frac{K}{2} \sin(x_l), \label{T2Limp}  \\
&&x_{l} = \lim_{n \rightarrow \infty}x_{2n}=\frac{p_l}{\Gamma(\alpha)}
\lim_{n \rightarrow \infty}\sum^{2n-1}_{k=0}(-1)^{k+1} V_{\alpha}^1(2n-i) 
=\frac{p_l}{\Gamma(\alpha)}V_{\alpha l}(k) \label{T2Limx},
\end{eqnarray}
where 
\begin{equation} \label{Valpl} 
 V_{\alpha l}  =  \sum_{k=1}^{\infty} (-1)^{k+1} V_{\alpha}^1(k).
\end{equation} 
Finally, the equation for the $x_l$ takes the form
\begin{equation}  \label{T2AntiXleq} 
x_l = \frac{K}{2 \Gamma(\alpha)} V_{\alpha l} \sin(x_l).
\end{equation} 
The numerical solution of Eqs.~(\ref{T2AntiXleq})~and~(\ref{T2Limp})
for $K=4.5$ when $1<\alpha<2$ 
is presented in Figs.~\ref{T2sink}~(b)-(d).  Figs.~\ref{T2sink}~(e)~and~(f)
show how well  this solution agrees with the results of  
numerical simulations of individual trajectories. 
After 1000 iterations presented in Figs.~\ref{T2sink}~(e)~and~(f)  
the values of deviations  
$|p_n-p_l|$ and  $|x_n-x_l|$ are less than $10^{-7}$.

The condition of the existence of a solution for 
Eq.~(\ref{T2AntiXleq}) 
\begin{equation}\ \label{Kc1} 
K > K_{c1}(\alpha) = \frac{2 \Gamma(\alpha)}{V_{\alpha l}}
\end{equation}
is the condition of the existence of the antisymmetric $T=2$ sink.
This sink exists above the curve $K = K_{c1}$ on Fig.~\ref{figBif}.
For $\alpha=2$  Eq.~(\ref{Kc1}) 
produces the standard map condition $K>4$ (see Sec~\ref{2DStLog})
and for $\alpha=1$ it gives $K>2$ (see Sec~\ref{1DStLog}). 

\subsubsection{ Fixed Points }
\label{Fixed}

Numerical simulations show that as in the 1D and 2D cases, in the
case of fractional maps with $1<\alpha<2$ the condition of
the appearance of $T=2$ trajectories coincides with the condition
of the disappearance of the stable fixed point.
This result for the fractional standard map was demonstrated in
\cite{ME4} and for the fractional logistic map was submitted for
publication. Below we present two ways in which stability of 
the RL-standard map's $(0,0)$ fixed point can investigated.

In the vicinity of the fixed point $(0,0)$  the equation for the
deviation of a trajectory from the fixed point can be written as
\begin{eqnarray}
&& \delta p_{n+1} = \delta p_n - K \delta x_n , \label{00fixP} \\
&&\delta x_{n+1} = \frac{1}{\Gamma (\alpha )} 
\sum_{i=0}^{n} \delta p_{i+1}V_{\alpha}(n-i+1) .\label{00fixX}
\end{eqnarray}
\begin{figure}
\centering
\includegraphics[width=0.89\textwidth]{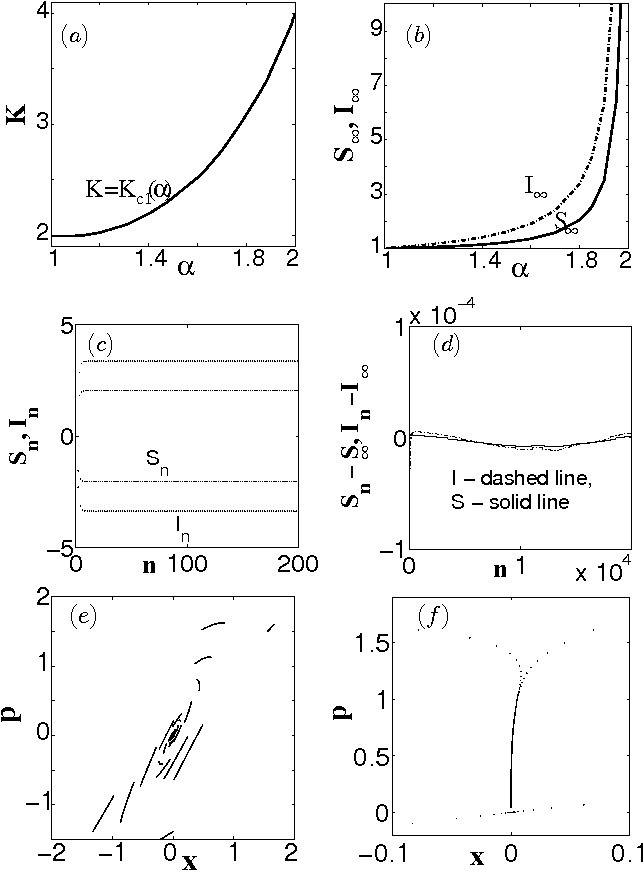}
\caption{\label{FigFix}  Stability of the fixed point $(0,0)$
in the RL-standard map with $1<\alpha<2$: (a). The fixed
point is stable below the curve $K=K_c(\alpha )$. (b). Values of $S_\infty$ 
and $I_\infty$ obtained after 20000 iterations of Eq.~\ref{SI}. 
The values of $S_\infty$ 
and $I_\infty$ increase rapidly when $\alpha \rightarrow 2$; for example, 
$S_\infty \approx 276$ and $I_\infty \approx 552$ 
after 20000 iterations when  $\alpha=1.999$. 
(c). An example of the typical evolution of $S_\infty$ and 
$I_\infty$ over the first 200 iterations for $1 < \alpha <2$.  
This particular figure corresponds to $\alpha =1.8$.
(d). Deviation of the values $S_n$ and $I_n$ from the values 
$S_\infty \approx 2.04337$ and  $I_\infty \approx 3.37416$ for 
$\alpha =1.8$ during the first 20000 iterations (this type of behavior 
remains for  $1 < \alpha <2$).
(e). Evolution of
trajectories with $p_0=1.5+0.0005i$, $0 \le i < 200$ for the case $K=3$,
$\alpha=1.9$. The line segments correspond to the $n$th iteration on the
set of trajectories with close initial conditions. 
The evolution of the trajectories with smaller $p_0$ is 
similar. (f). $10^5$ iterations on both of two trajectories for 
$K=2$, $\alpha =1.4$. The one at the bottom with $p_0=0.3$ is a fast
converging trajectory. The upper trajectory with  $p_0=5.3$ is an example
of an attracting slow converging trajectory in which $p_{100000} 
\approx 0.042$.   }
\end{figure}
Based on the results of Sec.~\ref{T2}  let's look for a solution in the form 
\begin{eqnarray}
&&\delta p_{n} = p_0\sum_{i=0}^{n-1}p_{n,i}\Bigl(\frac{2}{V_{\alpha
l}}\Bigr)^i\Bigl(\frac{K}
{K_{c1}(\alpha)}\Bigr)^i, \quad ( n > 0) , \label{00fixPSol} \\
&&\delta x_{n} = \frac{p_0}{\Gamma (\alpha )}\sum_{i=0}^{n-1}x_{n,i}
\Bigl(\frac{2}{V_{\alpha
    l}}\Bigr)^i\Bigl(\frac{K}{K_{c1}(\alpha)}\Bigr)^i, \quad (n > 0),  
\label{00fixXSol}
\end{eqnarray}
where $p_{n,i}$ and  $x_{n,i}$ satisfy the following iterative
equations
\begin{eqnarray}
&&x_{n+1,i}=-\sum_{m=i}^{n}(n-m+1)^{\alpha-1}x_{m,i-1} , \quad ( 0 <i \le n) ,
 \label{00fixXIter} \\ 
&&p_{n+1,i}=-\sum_{m=i}^{n}x_{m,i-1} , \quad ( 0 <i < n),  
 \label{00fixPIter} 
\end{eqnarray}
for which the initial and boundary conditions are
\begin{equation} \label{IC}
p_{n+1,n}=x_{n+1,n}=(-1)^n, \quad p_{n+1,0}=1, \quad x_{n+1,0}=(n+1)^{\alpha-1}.
\end{equation}
To verify the convergence of the alternating series 
Eqs.~(\ref{00fixPSol})~and~(\ref{00fixXSol}) we apply the Dirichlet's test
by considering the totals
\begin{equation}  \label{Dirichlet}
S_n=\sum_{i=0}^{n-1}x_{n,i}\Bigl(\frac{2}{V_{\alpha l}}\Bigr)^i, \quad
I_n=\sum_{i=0}^{n-1}p_{n,i}\Bigl(\frac{2}{V_{\alpha l}}\Bigr)^i.
\end{equation}
They obey the following iterative rules
\begin{equation} \label{SI}
S_n= n^{\alpha -1}-\frac{2}{V_{\alpha l}} \sum_{i=1}^{n-1}(n-i)^{\alpha -1}S_i,
\quad 
I_n=1 -\frac{2}{V_{\alpha l}} \sum_{i=1}^{n-1}S_i,
\end{equation}
where $S_1=1$.
Numerical simulations demonstrate that values of $S_n$ and $I_n$ converge 
to the values $(-1)^{n+1}S_\infty $ and $(-1)^{n+1}I_\infty $ presented 
in Fig.~\ref{FigFix}(b). Figs.~\ref{FigFix}(c)~and~(d) show an example of 
the typical evolution of  $S_n$ and $I_n$
over the first 20000 iterations. 
There is still no strict mathematical proof of the convergence.
From the boundedness of  $S_n$ and $I_n$ the convergence 
of $\delta p_{n}$ and  $\delta x_{n}$ requires the following condition
\begin{equation} \label{fixStable}
\frac{K}{K_{c1}(\alpha )} <1,
\end{equation}
which, as we expected, is exactly opposite to the condition of the
existence of the antisymmetric $T=2$ sink Eq.~(\ref{Kc1}).
Hundreds of runs of computer simulations confirmed that the transition
from the stable fixed point $(0,0)$ to the stable antisymmetric $T=2$
sink in both the RL-standard map and the Caputo-standard map occurs
on the curve $K=K_{c1}$ depicted in Fig.~\ref{FigFix}(a).

The second way to investigate stability of the $(0,0)$ fixed point is
by using generating functions \cite{Fel}, which in the case of
convolutions  
allows transformations of sums of products into products of sums.
After the introduction
\begin{equation}
\tilde{W}_{\alpha}(t)= \frac{K}{\Gamma (\alpha)}
\sum_{i=0}^{\infty }[(i+1)^{\alpha-1}-i^{\alpha-1}]t^i, \quad 
\tilde{X}(t)=\sum_{i=0}^{\infty }\delta x_i t^i,
\quad 
\tilde{P}(t)=\sum_{i=0}^{\infty }\delta p_i t^i
\label{GF}
\end{equation}
system Eqs.~(\ref{00fixP})~and~(\ref{00fixX}) can be written as
\begin{eqnarray}
&&\tilde{X}(t)=\frac{p_0 \tilde{W}_{\alpha}(t)}{K}  
\frac{t}{1 - t \Bigl(1- \tilde{W}_{\alpha}(t) \Bigr)  }, \label{GFeqx} \\ 
&&\tilde{P}(t)=p_0 \frac{1+  \tilde{W}_{\alpha}(t) }
{ 1-  t \Bigl( 1- \tilde{W}_{\alpha}(t) \Bigr) } . \label{GFeqp}
\end{eqnarray}
We see that the original problem can be solved by investigating  the 
asymptotic behavior at $t=0$ of the derivatives of the analytic functions
$\tilde{X}(t)$ and $\tilde{P}(t)$. This is still a complex
unresolved problem.

When $K<K_{c1}$ and the fixed point is stable, in phase space it is surrounded
by a finite basin of attraction, whose width $w$ depends on the values of $K$
and $\alpha$. For example, for $K=3$ and   $\alpha=1.9$ the width of the 
basin of attraction is $1.6<w<1.7$. 
Numeric simulations of thousands of trajectories 
with $p_0 < 1.6$ performed by the authors of \cite{ME4}, 
of which only 200 (with $1.5< p_0 < 1.6$) are presented 
in Fig.~\ref{FigFix}(e), 
show only converging trajectories,
whereas among 50 trajectories with  $1.6<p_0<1.7$ in Fig.~\ref{Fig12}(a) 
there are trajectories 
converging to the fixed point as well as some trajectories converging to 
attracting slow diverging trajectories, whose properties will be
discussed in the following section (Sec.~\ref{SecAttractors}). 
Fig.~\ref{FigFix}(e) shows fast converging trajectories. 
In the case $K=2$ and  $\alpha=1.4$ in addition to the fast converging  
trajectories  and attracting slow diverging trajectories
there exist attracting slow converging trajectories  
(Fig.~\ref{FigFix}(f)).

\subsubsection{Attractors Below Cascade of Bifurcations Band }
\label{SecAttractors}

In the following most of the statements  are conjectures made on the basis of
the results of
numerical simulations performed for some values of parameters $K$ and $\alpha$
which then were verified for additional  parameter values.  
\begin{figure}
\centering
\includegraphics[width=0.85\textwidth]{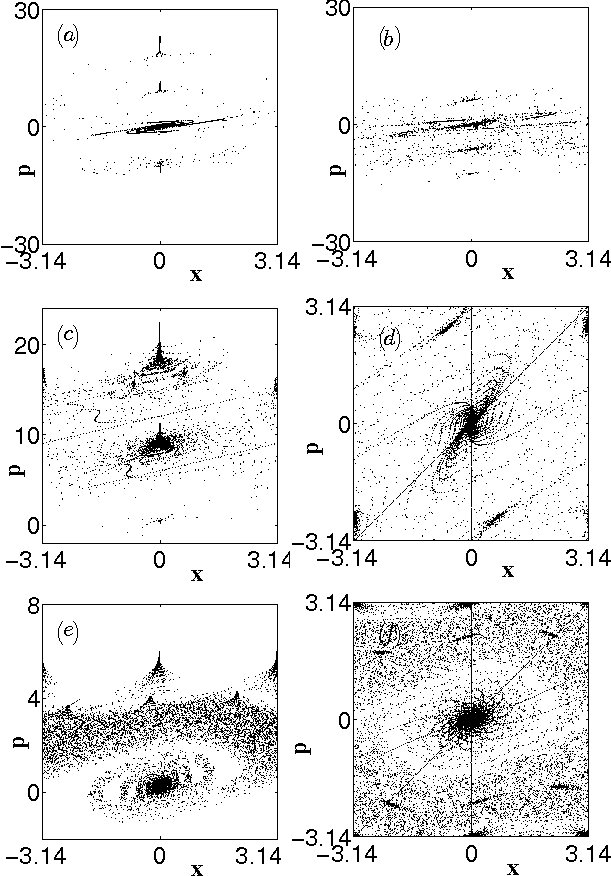}
\caption{\label{Fig12}  The  RL- and Caputo-standard maps' phase spaces 
for $K<K{c1}$: 
(a). The RL-standard map with the same values of parameters  as in
Fig.~\ref{FigFix}(e)  but  $p_0=1.6+0.002i$, $0 \le i < 50$.
(b). The Caputo-standard map with the same values of parameters  as in
Fig.~\ref{FigFix}(e)  but  $p_0=1.7+0.002i$, $0 \le i < 50$.
(c). 400 iterations on the RL-standard map trajectories with 
$p_0=4+0.08i$, $0 \le i < 125$ 
for the case  $K=2$, $\alpha=1.9$. 
Trajectories converge to the fixed point and 
two types of  attracting slow diverging
trajectories:  with $x_{lim}=0$ ($T=1$) and $T=4$.
(d). 100 iterations on the Caputo-standard map 
trajectories with $p_0=-3.14+0.0314i$, 
$0 \le i < 200$ for the same case as in (c) ($K=2$, $\alpha=1.9$)
but considered on a torus. 
In this case all trajectories converge to the fixed point or $T=4$ sink.
(e). 400 iterations on trajectories with $p_0=2+0.04i$, $0 \le i < 50$ 
for the RL-standard map case $K=0.6$, $\alpha=1.9$. 
Trajectories converge to the fixed point and two attracting slow diverging
trajectories ($T=2$ and $T=3$).   
(f).  100 iterations on the Caputo-standard map trajectories with $p_0=-3.14+0.0314i$, 
$0 \le i < 200$ for the same case as in (e) ($K=0.6$, $\alpha=1.9$)
considered on a torus. 
In this case all trajectories converge to the fixed point, period two and 
period three sinks. 
 }
\end{figure}

The structure of the fractional standard map's phase space
preserves some features which exist in the $\alpha = 2$ case.  
For example, for $K<K_{c1}$ stable higher period points, which exist in
the standard map, still exist in the fractional standard maps Fig.~\ref{Fig12}, 
but they exist in the asymptotic sense and
they transform from elliptic points into sinks and 
(in the case of the RL-standard map) into attracting slow 
($p_n \sim n^{2-\alpha}$) diverging trajectories.
In the area preserving standard map stable fixed and periodic points
are surrounded by  islands of regular motion which in the case of
fractional maps  turn 
into basins of attraction associated with sinks or 
slowly diverging attracting trajectories.
In the standard map islands are surrounded by chaotic areas.
For $K<K_{c1}$ and 
$1< \alpha <2$ in the fractional standard maps there are 
no chaotic or regular trajectories. Chaos exists in the following sense:
two initially 
close trajectories that start in an area between   
basins of attractions at first diverge,
but then converge to the same or different attractors.

\begin{figure}
\centering
\includegraphics[width=1.0\textwidth]{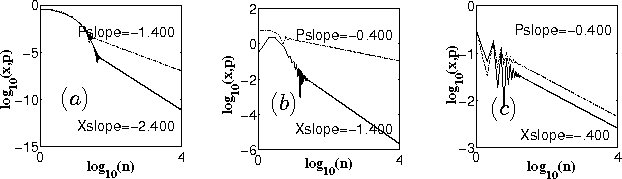}
\caption{\label{Fig13}  Different types of convergence of trajectories 
to the fixed point in the RL-standard map ((a) and (b)) and the
Caputo-standard map (c):
(a). Time dependence of the coordinate and momentum for the fast converging 
trajectory with $K=2$,  $\alpha =1.4$ and the initial conditions $x_0=0$ and 
$p_0=0.3$ from Fig.~\ref{FigFix}(f). 
(b). The same as in (a) but for the attracting slow converging 
trajectory with the initial conditions $x_0=0$ and 
$p_0=5.3$.
(c). $x$ and $p$ time dependence for the Caputo-standard map 
with $K=2$, $\alpha =1.4$, and the initial conditions  $x_0=0$ and $p_0=0.3$.  
 }
\end{figure}
There are differences not only between properties of the
regular and fractional standard maps but also between phase space
structures of the RL- and Caputo-standard maps. 
There is more than one way to approach an attracting periodic or fixed point of
the RL-standard map. 
In Fig.~\ref{Fig13} the examples of three trajectories, two
for the RL-standard map and one for the Caputo-standard map,
are used to demonstrate the differences in the rates of convergence.
In the RL-standard map
trajectories starting from  attractors' basins of attractions demonstrate
fast convergence with 
\begin{equation} \label{FastConv} 
\delta x_n \sim n^{-1-\alpha}, \  \ \delta p_n
\sim n^{-\alpha}
\end{equation}
and trajectories with the initial conditions from chaotic areas 
demonstrate slow convergence:  
\begin{equation} \label{SlowConv}
\delta x_n \sim n^{-\alpha}, \  \  \delta p_n
\sim n^{1-\alpha}.
\end{equation} 
There is only one type of convergence in the Caputo-standard map:
\begin{equation} \label{CaputoConv}
\delta x_n \sim n^{1-\alpha}, \  \  \delta p_n
\sim n^{1-\alpha}. 
\end{equation}
The same rates of convergence were observed also for antisimmetric 
(see Sec.~\ref{T2} and Fig.~\ref{Fig15})
and   $x_{n+1}=x_n-\pi$, $p_{n+1}=-p_n$  period two ($T=2$) points 
(Fig.~\ref{Fig16}).

From Figs.~\ref{Fig15}~(a)~and~(b) 
one can see that phase portraits on cylinders 
of the fractional standard maps with   $K=3$ and   $\alpha=1.9$ 
contain, in addition to the $(0,0)$ fixed point,
attracting slow diverging trajectories (RL-case), or fixed points
(Caputo-case) approximately equally spaced along the $p$-axis. 
This result agrees with  the fact that the standard map with  $K=3$ has only 
one central island. 
More complex structures of the fractional standard maps' phase spaces, for
$K=2$ with  $T=4$  sinks (Figs.~\ref{Fig15}~(c)~and~(d)) 
and for $K=0.6$ with  $T=2$ and $T=3$ sinks (Figs.~\ref{Fig15}~(e)~and~(f)), 
can be explained by the presence of the islands with the same periodicity
in the standard map with the same $K$.
\begin{figure}
\centering
\includegraphics[width=1.0\textwidth]{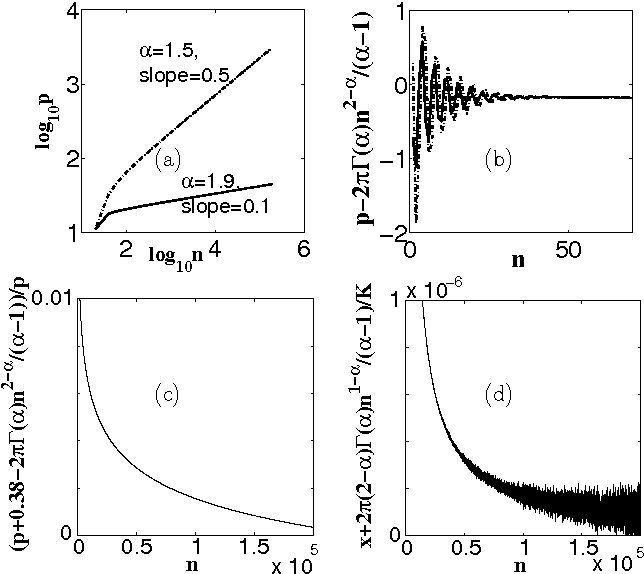}
\caption{\label{Fig14} Evaluation of the behavior of the attracting slow
diverging trajectories: 
(a). Momenta for two trajectories with $x_n \approx 2\pi n$
in  unbounded space (in this example $K=2$). 
The solid line is related to a trajectory with 
$\alpha = 1.9$ and its slope is 0.1. The dashed line  corresponds to a 
trajectory with $\alpha = 1.5$ and its slope is 0.5.
(b). Deviation of momenta from the asymptotic formula 
for two trajectories with $x_n \approx 2\pi n$
in  unbounded space, $\alpha = 1.9$, and  $K=2$. The dashed line 
has $p_0=7$ and the solid one $p_0=6$. 
(c). Relative deviation of the momenta for
the trajectories in (b) from the asymptotic formula.
(d).  Deviation of the $x$-coordinates  for
the trajectories in (b) from the asymptotic formula.
 }
\end{figure}
Numerical evaluations  (see Fig.~\ref{Fig14}) lead to the suggestion 
that attracting slow
diverging trajectories which converge to 
trajectories along the $p$-axis ($x \rightarrow x_{lim}=0$) 
in the area of parameters of their stability
for large $n$ demonstrate
the following asymptotic behavior 
\begin{equation} \label{5i}
p_n = C n^{2-\alpha }.
\end{equation}
The constant C can be evaluated for $1.8< \alpha <2$. 
Consider a trajectory on a cylinder with $ x_{lim}=0$, $T=1$, and 
constant step in $x$ in the unbounded space $2 \pi M$,
where $M$ is an integer.
Then from Eq. (\ref{FSMRLx}) follows
\begin{equation} \label{5}
x_{n+1}-x_{n} = \frac{1}{\Gamma (\alpha )} 
\sum_{k=1}^{n} (p_{k+1}-p_k)V_{\alpha}^1(n-k+1)
+ \frac{p_1}{\Gamma (\alpha)} V_{\alpha}^1(n+1) .
\end{equation}
For large $n$ the last term is small ($\sim n^{\alpha-2}$) and the 
following holds
\begin{equation} \label{6} 
\sum_{k=1}^{n} (p_{k+1}-p_k)V_{\alpha}^1(n-k+1) = 2 \pi M \Gamma (\alpha).
\end{equation}

It can be shown, assuming $p_n \sim  n^{2-\alpha}$,
that for values of $\alpha>1.8$ the terms in the last
sum with large $k$ are small and in the series representation
of $V_{\alpha}^1(n-k+1)$ only terms of the 
highest order in $k/n$ can be kept.
In this case, Eq.~(\ref{6}) leads to the approximations 
\begin{equation} \label{7} 
p_n \approx p_0 +  \frac{2 \pi M \Gamma (\alpha) n^{2-\alpha}}{\alpha-1},
 \quad 
x_n \approx -\frac{2\pi M(2-\alpha ) \Gamma (\alpha) }{ K(\alpha-1) n^{\alpha-1}}.
\end{equation}
In the case $K=2$, $\alpha=1.9$ Figs.~\ref{Fig14}~(b)-(d) 
show for two  trajectories with $M=1$
(initial momenta $p_0=6$ and  $p_0=7$) approaching an  attracting slow
diverging trajectory the deviation
from the asymptotic formula Eq.~(\ref{7})  
and the relative difference with respect to Eq.~(\ref{7}).

\begin{figure}
\centering
\includegraphics[width=1.0\textwidth]{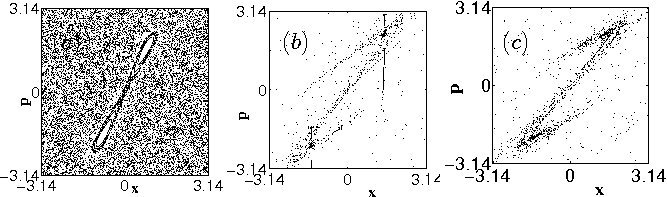}
\caption{\label{Fig15}Stable antisymmetric 
$x_{n+1}=-x_n$,  $p_{n+1}=-p_n$ period   $T=2$ trajectories for
$K=4.5$: (a). 1000 iterations on each of 25 trajectories 
for the standard map with $K=4.5$.  The only feature is a system of two
islands associated with the period two elliptic point.
(b). RL-standard map stable $T=2$ antisymmetric sink 
for $\alpha=1.8$.  500 iterations on each of 25 trajectories: 
$p_0=0.0001+0.08i$, $0 \le i <25$. Slow and fast converging trajectories.
(c). Caputo-standard map stable $T=2$ antisymmetric sink 
for $\alpha=1.8$. 1000 iterations on each of 10 trajectories: 
$p_0=-3.1415+0.628i$, $0 \le i <10$.  
 }
\end{figure}
As for $K<K_{c1}$, in the case  $K_{c1}(\alpha)<K<K_{c2}(\alpha)$ 
asymptotic existence and stability
of the antisymmetric sink (Sec.~\ref{T2}) 
is a result of the gradual transformation of the standard map's
elliptic point with the decrease in the
order of derivative from  $\alpha =2$  
(see Fig.~\ref{Fig15}). Convergence of trajectories 
follows Eqs.~(\ref{FastConv})-(\ref{CaputoConv}).

The standard map's antisymmetric $T=2$ trajectory 
becomes unstable when $K=2\pi$ and at the point $(\pi/2,0)$
in phase space a pair of $T=2$ trajectories with 
$x_{n+1}=x_n-\pi$,  $p_{n+1}=-p_n$ appears. Numerical simulations 
of the fractional standard maps (see Fig.~\ref{Fig16})
show that they demonstrate similar behavior.
With the assumption that the RL-standard map 
Eqs.~(\ref{FSMRLp})~and~(\ref{FSMRLx}) 
have an asymptotic solution
\begin{equation} \label{T2nonAS} 
p_{n} = (-1)^np_l, \    \  x_{n} = x_l-\frac{\pi}{2}[1-(-1)^n]
\end{equation}
it can be shown  from Eq.~(\ref{FSMRLp}) that the relationship  
$p_l = K/2 \sin(x_l)$ (Eq.~(\ref{T2Limp}))
is valid in this case too.

Numerical simulations similar to those presented in Fig.~\ref{Fig13} 
show that for $K>K_{c2}$ (see Fig.~\ref{figBif}) the RL-standard map 
has the asymptotic behavior
\begin{equation} \label{T2nonLimAS} 
p_{n} = (-1)^np_l+An^{1-\alpha},
\end{equation}
where $A$ is the same for both even and odd values of $n$.
\begin{figure}
\centering
\includegraphics[width=1.0\textwidth]{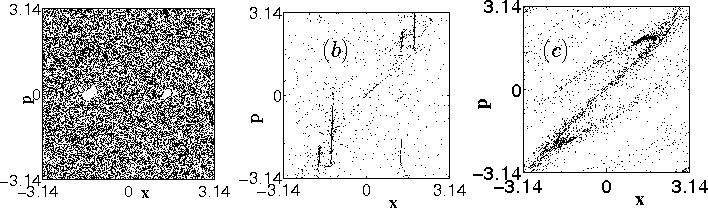}
\caption{\label{Fig16} Stable  
$x_{n+1}=x_n-\pi$,  $p_{n+1}=-p_n$ period   $T=2$ trajectories for
$K>K_{c2}$: (a). 500 iterations on each of 50 trajectories 
for the standard map with $K=6.4$.  The main features are two accelerator mode 
sticky islands around points $(-1.379,0)$ and $(1.379,0)$ which define 
the dynamics. Additional features - dark spots at the top and the bottom
of the figure (which are clear on a zoom) - 
two systems of $T=2$ tiny islands 
associated with two  $T=2$ elliptic points:  $(1.379,\pi)$,
$(1.379-\pi,-\pi)$ and $(\pi-1.379,\pi)$, $(-1.379,-\pi)$.
(b). Two RL-standard map's stable $T=2$ sinks for $K=4.5$, $\alpha=1.71$. 
500 iterations on each of 25 trajectories: 
$p_0=0.0001+0.08i$, $0 \le i <25$.
(c). Two Caputo-standard map's stable $T=2$ sinks 
for  $K=4.5$, $\alpha=1.71$. 1000 iterations on each of 10 trajectories: 
$p_0=-3.1415+0.628i$, $0 \le i <10$.  
 }
\end{figure}
After substituting (\ref{T2nonLimAS}) in (\ref{FSMRLx}) in the  
limit $n \rightarrow \infty$ one can derive 
\begin{equation} \label{T2nonASxsol} 
\sin(x_l)= \frac{\pi \Gamma(\alpha)}{K V_{\alpha l}}, 
\end{equation}
which has solutions when
\begin{equation} \label{T2nonASKc} 
K>K_{c2}= \frac{\pi \Gamma(\alpha)}{V_{\alpha l}} 
\end{equation}
(see Fig.~\ref{figBif}). The  value of $A$ can also be calculated:
\begin{equation} \label{T2nonASA} 
A= \frac{2 x_l-\pi}{2 \Gamma(2-\alpha)}. 
\end{equation}
Results of the analytic estimations
Eqs.~(\ref{T2nonASxsol})~-(\ref{T2nonASA}) 
are in  good agreement with the direct numerical
simulations of the fractional standard maps.

\subsubsection{Cascade of Bifurcations Band }
\label{BisBand}

\begin{figure}
\begin{center}
\includegraphics[width=1.\textwidth]{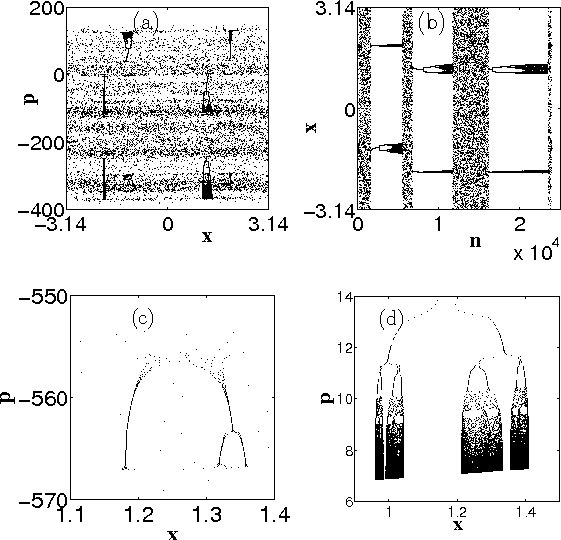}
\caption{
Cascade of bifurcations type trajectories in the RL-standard map:
(a). $\alpha=1.65, K=4.5$; one intermittent trajectory in phase space.
(b). Time dependence of the coordinate $x$ ($x$ of $n$) for the case (a).
(c). $\alpha=1.98, K=6.46$; zoom of a small feature for a single intermittent 
trajectory in phase space.
(d). $\alpha=1.1, K=3.5$; a single trajectory enters the cascade after 
a few iterations and stays there 
during 500000 iterations. 
}
\label{Cascades}
\end{center}
\end{figure}

At $K \approx 6.59$ in the standard map $T = 2$ points become unstable
and stable $T = 4$ elliptic points appear. 
Further increase in $K$ results in the period doubling cascade of 
bifurcations which leads to the disappearance of
the corresponding islands of stability in the chaotic 
sea at $K \approx 6.6344$ (see Sec.~\ref{2DStLog}).
The cusp in Fig.~\ref{figBif}(a) points to a point $\alpha = 2$ and 
$6.59<K_*<6.63$.
Inside the
band leading to the cusp a new type of attractors, cascade of
bifurcations type trajectories (CBTT), appears (see Fig.~\ref{Cascades}).
The lower boundary of the band approximately corresponds to the 
transition from the $T=2$ sink $x_{n+1}=x_n-\pi$, $p_{n+1}=-p_n$ 
to the $T=4$ sink and the upper boundary 
corresponds to the transition to chaos.  
At $\alpha=1$ the lower and upper boundaries correspond to  the
$T=2$ $\rightarrow$ $T=4$ transition and the transition to chaos in
the 1D standard map (see Sec.~\ref{1DStLog}).
In  CBTT period doubling
cascade of bifurcations occurs on a single trajectory with a fixed 
value of the map parameter. 
A typical CBTT's behavior is similar to the
behavior of trajectories in Hamiltonian
dynamics in the presence of sticky islands: occasionally 
a trajectory enters a CBTT and then leaves it and enters
the chaotic sea (Figs.~\ref{Cascades}~(a)~and~(b)).  
With the  decreases in  $\alpha$ the relative time
trajectories spend in CBTT increases.
CBTT are barely distinguishable near the cusp (Fig.~\ref{Cascades}(c))
and trajectories spend relatively little time in CBTT.
A trajectory enters
a CBTT after a few iterations and stays there over the longest computational 
time we were running our codes - 500000 iterations when $\alpha$ is 
close to one.

The CBTT in Fig.~\ref{Cascades} were obtained for the RL-standard map.
In many cases it is difficult to find CBTT  
in phase space of the Caputo-standard map but 
they look almost the same for both fractional maps
on the $x$ vs. $n$
plot (see Fig.~\ref{Cascades}(b)).

Results of numerical simulations submitted for publication 
show that not CBTT but inverse (in time) CBTT,  are present within the 
CBTT band (from the $T=2$ $\rightarrow$ $T=4$
transition to the transition to chaos) 
of the fractional logistic maps.

\subsubsection{More Fractional Attractors}
\label{MoreAttractors}

\begin{figure}
\centering
\includegraphics[width=1.\textwidth]{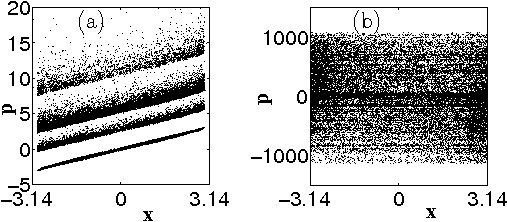}
\caption{\label{proper} ``Proper'' and ``improper'' attractors in
the  RL-standard map. 
3000 iterations on ten 
trajectories with the initial conditions 
$x_0=0$, $p_0=0.001+1.65i$, $i=0,1,...9$:
(a). A ``proper'' chaotic attractor for $K=4.2$, $\alpha=1.1$.
(b). An ``improper'' chaotic attractor for $K=4.4$, $\alpha=1.1$.
}
\end{figure}

In the one-dimensional standard map with $K>0$ the 
``proper'' chaotic attractor 
exists for $3.532<K< 4.603339$ (see Sec.~\ref{1DStLog}). This is the
interval between the upper boundary of the CBTT band for $\alpha=1$ and
$K=K_{c3}(1)$ in Fig.~\ref{figBif}.
In the area between   $K=K_{c3}(\alpha)$ curve and the upper border of the 
CBTT band (in Fig.~\ref{figBif}) 
the fractional chaotic attractors are proper (see Fig.~\ref{proper}(a)) and
above   $K=K_{c3}(\alpha)$ the entire phase space is  chaotic 
(Fig.~\ref{proper}(b)).

The standard map has a set of bands for $K$ above $2\pi n$ 
of the accelerator mode sticky islands in which momentum increases
proportionally to the number of iterations $n$ and coordinate increases as
$n^2$ (see Sec.~\ref{2DStLog}). 
In the one-dimensional standard map the corresponding bands demonstrate  cascades of
bifurcations (see Fig.~\ref{figBif}(b)) for $|K|$ above $2\pi |n| $. 
The acceleration in those bands is zero and $x$ increases
proportionally to $n$ (see Sec.~\ref{1DStLog}).
\begin{figure}
\centering
\includegraphics[width=1.\textwidth]{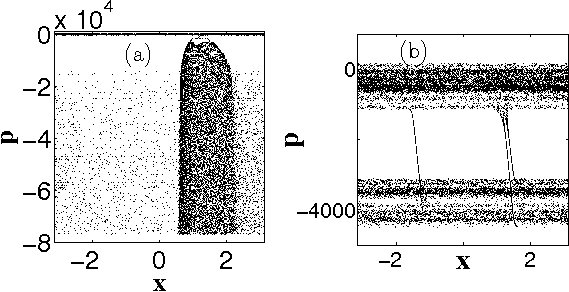}
\caption{\label{Ballistic} RL-standard map's accelerator mode attractors. 
25000 iterations on a
 single trajectory with the initial conditions $x_0=0$, $p_0=0.1$:
(a). CBTT-type  accelerator mode attractor for $K=5.7$, $\alpha=1.03$.
(b). Accelerator mode attractor for $K=7.6$, $\alpha=1.97$.
}
\end{figure} 

Accelerator mode attractors in the 
case $1<\alpha<2$ are not fully investigated. The standard map's accelerator
mode islands evolve into the accelerator mode (ballistic) attracting sticky 
trajectories when $\alpha$ is reduced from $2$ for the values of $K$ which
increase with the decrease in $\alpha$ (Fig.~\ref{Ballistic}(b)). When the  
value of $\alpha$ increases from 1, the corresponding ballistic attractors
evolve into the cascade of bifurcation type ballistic trajectories 
(see Fig.~\ref{Ballistic}(a)) for the values of  $K$ which
decrease with the increase in $\alpha$. This could mean that corresponding 
features in the one- and two-dimensional maps (at least for $K=2\pi$) 
are not connected by the continuous change in $\alpha$.

\subsection{$\alpha$-Families of Maps ($2<\alpha<3$)}
\label{BN23}

Fractional maps for $\alpha>2$ are not yet investigated. Here we'll
present the first results \cite{DNC} for the RL-standard map.

\begin{figure}
\centering
\includegraphics[width=1.\textwidth]{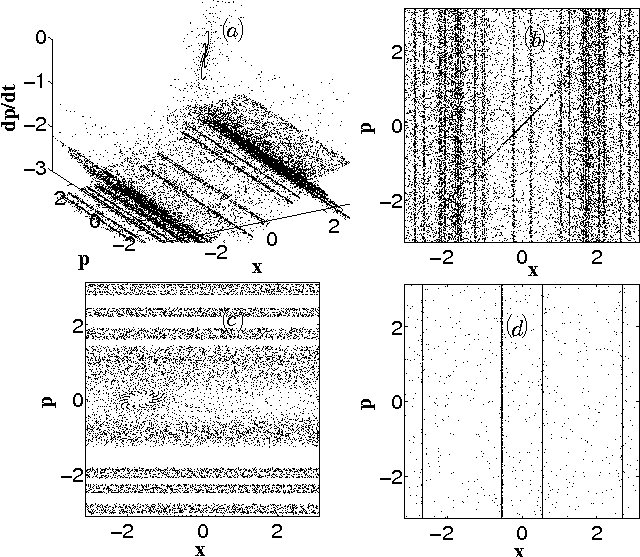}
\caption{\label{RL2_3D} RL-standard map for $2< \alpha <3$:
(a). 3D phase space for $K=1$, $\alpha=2.01$ obtained on a single trajectory
with $x_0=p_0=0$ and $p^1_0=0.01$. (b). Projection of the phase space in (a)
on the $x$-$y$ plane. (c). Projection of the phase space for  $K=0.2$, 
$\alpha=2.01$, $x_0=p_0=0$  
on the $x$-$y$ plane obtained using 20 trajectories with different 
initial values  of $p^1_0$. (d). The same as in (c), but for  $K=4$ and 
$\alpha=2.9$.   
}
\end{figure}
With $G_K(x)=K \sin (x)$  in Eqs.~(\ref{FrRLMapx})~and~(\ref{FrRLMapp}), 
the RL-standard map for 
$2< \alpha \le 3$ can be written as
\begin{eqnarray}
&&p^1_{n+1}= p^1_n-K\sin(x_n), \nonumber 
\label{SMRLalp2n3p1} \\
&&p_{n+1}= p^1_n +p_n-K\sin(x_n), \   \  ({\rm mod} \ 2\pi ),
\label{SMRLMalp2n3p} \\
&&x_{n+1}=\frac{p_0}{\Gamma(\alpha-1)}(n+1)^{\alpha-2}+\frac{1}{\Gamma(\alpha)}\sum^{n}_{k=0}
p^1_{k+1}V^1_{\alpha}(n-k+1), \   \ ({\rm mod} \ 2\pi ). \nonumber 
\label{SMRLalp2n3x} 
\end{eqnarray} 
In our simulations we did not find a stable fixed point even for small
values of $K$ (see Fig.~\ref{RL2_3D}~(c)). 
Simulations show that for this map there are attractors in the form of 
the attracting multi-period lines with constant $x$ (see
Fig.~\ref{RL2_3D} (a), (b), and (d)). For most of the values of the map 
parameters the phase space is highly chaotic.  
   
This case and the transition from the 2D standard map to the 3D standard
map is not yet fully investigated.

\section{Conclusion}
\label{Conclusion}

The systems with long-term memory that are most frequently encountered 
in nature are systems with power-law memory.
In many applications, 
including biological applications, the exponent in power law,
 $\sim t^{-\beta}$, is $0<\beta<1$. This is true, in particular, for
adaptive systems and for viscoelastic properties of human tissues.
These systems can be described by nonlinear fractional differential equations
with fractional derivatives of the order $\alpha=1-\beta$ with
$0<\alpha<1$. Fractional differential equations can be modeled by 
discrete nonlinear maps with power-law memory. We studied  maps
which model fractional differential equations with $0<\alpha<2$
and, correspondingly, $-1<\beta<1$. 
Decrease in $\beta$ and,  correspondingly, increase in $\alpha$  
means an increase in the memory effects - older states have higher
weights in the definition of the present state of a system. 

In Sec.~\ref{AFM} we showed that an increase in memory effects leads 
to more complicated and chaotic behavior. As can be seen in
Fig.~\ref{LowAlpBif}, systems with small $\alpha$ are more stable. 
At the values of system parameters, corresponding to the
periodic behavior and transition to chaos, behavior of such systems
follows a well defined cascade of bifurcations pattern  Fig.~\ref{CBTT1D}.
This type of evolution may  mean a slow adaptation
when a system changes its state  long after a change in a parameter
occurred. 

Increase in memory effects with the transition from $0<\alpha<1$ to  
$1<\alpha<2$ leads to increased diversity in systems' behavior.
Systems with $1<\alpha<2$  may demonstrate
periodic sinks, attracting slow diverging trajectories,
attracting accelerator mode trajectories, 
chaotic attractors, and cascade of
bifurcations and inverse cascade of bifurcations type attracting trajectories. 
An intermittent cascade of bifurcations type behavior 
(Figs.~\ref{Cascades}~(a)~and~(b))  may correspond to 
a scenario of the evolution of chronic diseases, to some mental
disorders, or to the evolution of some social systems.  

The way in which systems with power-law memory 
approach fixed and periodic points
(Eqs.~(\ref{FastConv})-(\ref{CaputoConv}))
can be used to identify systems with memory in  an analysis 
of experimental data.

\section*{Acknowledgments}
The author expresses his gratitude to V.~E. Tarasov for useful
discussions, 
to E. Hameiri and H. Weitzner 
for the opportunity to complete this work at the Courant Institute,
and to 
V. Donnelly for  technical help.

\bibliographystyle{spphys} 
\input{MEreferencOrdered}

\end{document}

%% file: MEreferencOrdered.tex
%
%
%